\documentclass[backref, breaklinks,colorlinks,twocolumn]{aastex701}
\hypersetup{linkcolor=blue,citecolor=blue,filecolor=cyan,urlcolor=magenta}
\usepackage{graphicx}
\usepackage{ulem}
\usepackage{epstopdf}
\usepackage{color}
\usepackage{hyperref}
\usepackage{times}
\usepackage[utf8]{inputenc} 
\usepackage{float}
\usepackage{bm}
\usepackage{ulem}
\usepackage{multirow}
\usepackage{makecell}
\usepackage{url}
\usepackage{mathrsfs}
\usepackage{physics}
\usepackage{comment}

\newcommand{\dalm}{\kern1pt\vbox{\hrule height 0.9pt\hbox{\vrule width 0.9pt
\hskip 2.5pt\vbox{\vskip 5.5pt}\hskip 3pt\vrule width 0.3pt}\hrule height 0.3pt}
\kern1pt}

\begin{document}

\title{Estimation of neutron star mass and radius of FRB 20240114A by identification of crustal oscillations}




\author[0000-0002-3239-2921]{Hajime Sotani}
\email{sotani@yukawa.kyoto-u.ac.jp}
\affiliation{Department of Mathematics and Physics, Kochi University, Kochi, 780-8520, Japan}
\affiliation{RIKEN Center for Interdisciplinary Theoretical and Mathematical Sciences (iTHEMS), RIKEN, Wako 351-0198, Japan}
\affiliation{Theoretical Astrophysics, IAAT, University of T\"{u}bingen, 72076 T\"{u}bingen, Germany}

\author[0000-0002-9249-0515]{Z. Wadiasingh}
\affiliation{Department of Astronomy, University of Maryland, College Park, MD 20742-2421, USA}
\affiliation{Astrophysics Science Division, NASA Goddard Space Flight Center, Greenbelt, MD 20771, USA}
\affiliation{Center for Research and Exploration in Space Science and Technology, NASA/GSFC, Greenbelt, MD 20771, USA}
\email{zorawar.wadiasingh@nasa.gov}

\author[0000-0003-2759-1368]{C.~Chirenti}
\affiliation{Department of Astronomy, University of Maryland, College Park, MD 20742-2421, USA}
\affiliation{Astrophysics Science Division, NASA Goddard Space Flight Center, Greenbelt, MD 20771, USA}
\affiliation{Center for Research and Exploration in Space Science and Technology, NASA/GSFC, Greenbelt, MD 20771, USA}
\email{chirenti@umd.edu}




\begin{abstract}
By identifying quasi-periodic oscillations (QPOs) reported in FRB 20240114A (from the Five-hundred-meter Aperture Spherical Telescope) with neutron star crustal torsional oscillations, together with experimental constraints on the incompressibility $K_0$ of symmetric nuclear matter at saturation density, we constrain the mass and radius of an extragalactic neutron star at redshift $z\approx0.13$. Identifying the low-order QPO frequencies as fundamental oscillations, and frequencies of $567.7\,\mathrm{Hz}$ or $655.5\,\mathrm{Hz}$ (rest frame) as first overtone candidates, implies neutron star mass ranges of $1.00$--$1.55\,M_\odot$ or $1.17$--$1.76\,M_\odot$, respectively. The radius is also constrained, with a self-consistent value around $13$~km, consistent with the calculation of the NS structure within the low-mass/low-central density regime. Simultaneously, we also constrain another nuclear saturation parameter, namely the density dependence of the nuclear symmetry energy at saturation density (i.e., the slope parameter), $L$, and determine it to be $L=59.5-96.8$ MeV with $\sim 10\%$ systematic uncertainty, which is broadly consistent with previous constraints on $L$ obtained from experiments and astronomical observations. Thus, a mapping of FRB QPOs to crustal torsional modes seems reasonable. This interpretation will be tested with the discovery of additional QPOs in upcoming FRB surveys.
\end{abstract}



\section{Introduction}
\label{sec:I}
\tighten


Fast radio bursts \citep[FRBs,][]{1980ApJ...236L.109L,doi:10.1126/science.1147532,doi:10.1126/science.1236789,2022MNRAS.515.3698C,2024Ap&SS.369...59L} are energetic radio phenomena with rest-frame durations ranging from microseconds to several milliseconds. Most are extragalactic, with activity levels ranging from single bursts to thousands of repeat bursts from an individual source \citep{2019A&ARv..27....4P,2021Univ....7..453C}. Their origins and mechanisms are still uncertain with many proposed models \citep[e.g.,][]{2019PhR...821....1P,2021SCPMA..6449501X,2021sf2a.conf..405V,2023RvMP...95c5005Z}. Magnetized neutron stars (NSs), particularly magnetars, are leading engine candidates. However, even magnetar models have a range of proposed emission locales, physical mechanisms, triggers, and source conditions. Most early models invoked magnetar giant flares with FRB generation and escape taking place at large distances in shocks \citep[e.g.,][]{Popov2010,2014MNRAS.442L...9L,2017ApJ...843L..26B,2019MNRAS.485.4091M,2020ApJ...896..142B}. However, shock models with `hyperactive' young magnetars suffer from low efficiency and inability to explain short timescale microstructure \citep{2020MNRAS.498..651B}, polarization jumps/swings \citep[e.g.,][]{2024ApJ...972L..20N,2025Natur.637...43M}, or scintillation constraints \citep{2025Natur.637...48N}. Magnetospheric magnetar models also range in physical conditions and mechanisms \citep[e.g.,][]{2017MNRAS.468.2726K,2018ApJ...852..140W,Wadiasingh2019,2019MNRAS.488.5887S,Wadiasingh2020,2021ApJ...922..166L,2022MNRAS.517.5080W}. As already proposed prior to 2020, FRBs likely result from crust quakes and perturbations associated with magnetar short bursts \citep{Wadiasingh2019,Wadiasingh2020} where not all short bursts result in FRBs due to a threshold condition for the amplitude of crust perturbations. This is also consistent with lower-energy breaks observed in numerous individual repeating FRB luminosity/energy distributions \citep[e.g.,][]{2025arXiv251221889K}, i.e., a preferred energy scale for FRBs, and the earthquake-like statistics of some FRBs \citep{2023MNRAS.526.2795T}.

The magnetar engine for (at least some) FRBs was also empirically corroborated by the galactic event FRB 20200428A and its associated soft gamma-ray short burst, from magnetar SGR 1935+2154, in 2020~\citep{2020Natur.587...54C,2020Natur.587...59B,2020ApJ...898L..29M,2021NatAs...5..372R,2021NatAs...5..378L,2021NatAs...5..401T}. This was preceded by thousands of magnetar short bursts with no observed FRB-like emission \citep{2020ApJ...904L..21Y}. 
More recently, during the outburst activity of SGR 1935+2154 in 2022 accompanied by short bursts, FRB-like bursts were temporally bracketed by two spin-up glitches \citep{2024Natur.626..500H}, the largest ever observed in NSs. This activity is also contemporaneous with enhanced spin-down, along with rapid X-ray spectral evolution \citep{2025ApJ...989...63H}, confirming some FRBs are the result of crustal activity with peculiar magnetospheric conditions. 

A few percent of observed (extragalactic) FRBs contain sub-bursts or multipulse trains of bursts, separated by characteristic timescales of a few to tens of milliseconds \citep[e.g.,][]{2019ApJ...876L..23H,2021ApJ...923....1P}. Additionally, repeater FRBs seem to universally feature a double-humped waiting time distribution \citep{2025ApJ...983L..16Y}, with no short-timescale periodicity \citep[e.g.,][]{2022RAA....22l4004N,2024MNRAS.52710425S,zhou2025comprehensivesearchlongshort,2025arXiv251224936K}. It was proposed in \citet{Wadiasingh2019} that the short-time hump is not simply microstructure but an imprint of engine-driven activity, i.e., the characteristic timescale of shear modes in the magnetar crust, and further expounded in \citet{Wadiasingh_2020}. Indeed, such a short-time hump is also seen in high-energy magnetar short bursts \citep{2015ApJ...810...66H} along with quasi-periodic oscillations (QPOs) compatible with low-order fundamental crustal eigenmodes \citep{2014ApJ...795..114H,2014ApJ...787..128H}. The lack of short timescale periodicity (associated with a more typical spin period of $1-10$~s) is consistent with the FRB engine being an ultra-long period magnetar \citep{Wadiasingh2020,Beniamini+20,2023MNRAS.520.1872B,2024MNRAS.533.2133C,2025A&A...696A.194B,2025ApJ...982...45B,2025arXiv250505373S,Lander_2026}. Long term polarimetric monitoring also seems to favor this picture \citep{2025A&A...702A.248B}.

Remarkably, the HXMT-Insight hard X-ray component of FRB 20200428A also exhibits a $\sim 3.4\sigma$ QPO at $\sim40$~Hz \citep{2022ApJ...931...56L}, with a period compatible with the interval between radio peaks in  FRB 20200428A. Thus, QPOs, if associated with crustal modes and perturbations, are likely capable of being imprinted on observed FRB interval times or substructure\footnote{The recent study by ~\citet{2025ApJ...995L..57B} considered simulations of large-amplitude MHD-like waves in a force-free approximation for the imprinting of crust modes that may steepen into shocks or interact with upstream plasma.  The background plasma (sourced by single-photon pair cascades) must exist to support such waves without charge starvation and/or being Compton thick, which is not a given, while also not inhibiting FRB escape. This contrasts with a low-twist magnetosphere with nearly vacuum-like conditions, where crust perturbations drive pair creation and radio emission with a threshold condition.
Furthermore, \cite{2026ApJ...998..190Q} recently studied the generation of MHD waves from seismic motions and studied their wave propagation in the crust, which might also be relevant.}. The interval spacing also disfavors strong beaming of emission, given the magnetar's spin. 
The FRB subpulses also led\footnote{See also \cite{wang2026gecamdiscoverysecondfrbassociated} for FRB 20221014 from SGR 1935+2154 with similar lag.} X-ray features by $\sim 2-3$~ms~\citep{2023arXiv231016932G,2023ApJ...953...67G}, which is compatible with a gap Spitzer timescale~\citep{2025ApJ...980..211B} and is unexplained in shock or triggered reconnection models. 

Thus, the close connection of FRB activity with NS crust physics, and identification\footnote{We note that the imprinting mechanism proposed by \citet{2025ApJ...995L..57B} arises from quake-driven elastic motions with a characteristic crust-crossing/bounce frequency (typically kHz in their model, well above low-order fundamental torsional oscillation frequencies), coupled into force-free MHD magnetospheric disturbances. Low-order fundamental oscillations or magneto-elastic modes, as exhibited by FRB 20200428A and its associated short burst, seem also capable of being imprinted \citep[][]{Wadiasingh_2020}. Our approach instead tests whether the reported FRB QPOs can be mapped onto torsional-mode eigenfrequencies in simplified (non-magnetic) crust models.} of subpulse trains in FRBs with possible coherent crustal modes rather than burst microstructure \citep[e.g.,][]{2024NatAs...8..230K} is supported by numerous pieces of evidence. We adopt this working hypothesis, which, as demonstrated in the paper, yields results intriguingly consistent with a typical NS. There are two important caveats to our work. First, we assume that the reported QPOs are real, even though their statistical significance is sometimes low. And second, we note that consistency is not enough to rule out alternatives to our hypothesis, but it can be tested against future (and imminent) large FRB samples from CHORD \citep{2019clrp.2020...28V} and DSA \citep{2019BAAS...51g.255H,2024AAS...24326104S}.

Examples of FRBs with reported QPO periods or clustering include FRB~20201020A \citep[0.41~ms,][]{2023A&A...678A.149P}, FRBs 20210206A and 20210213A \citep[2.8~ms and 10.7~ms, respectively,][]{2022Natur.607..256C}, FRB~20230708A \citep[7.3~ms,][]{2025MNRAS.536.3220D}, FRB~20220912A~\citep[5.8 ms,][]{2024MNRAS.52710425S}, FRB~20190122C \citep[1~ms,][]{xiao2026evidencedampedmillisecondquasiperiodic}, and FRB 20121102A \citep[22~ms,][]{2023MNRAS.519..666J}.

More recently, FRB 20240114A~\citep{2024ATel16420....1S} possesses an exceptionally large burst sample enabled by the Five-hundred-meter Aperture Spherical radio Telescope (FAST) \citep{2011IJMPD..20..989N,zhang2025investigatingfrb20240114afast} with numerous modest significance radio QPOs~\citep{zhou2025comprehensivesearchlongshort}. We choose to focus on this sample for this study. The reported QPO frequencies range from tens of Hz up to $\sim 600$ Hz, which are comparable to the QPOs discovered in X-rays from magnetar giant flares \citep[e.g.,][]{2005ApJ...632L.111S,2006ApJ...637L.117W,2006ApJ...653..593S,2007AdSpR..40.1446W,Miller_2019} and short bursts \citep{2014ApJ...795..114H,2014ApJ...787..128H}.

This manuscript is organized as follows. In Sec.~\ref{sec:modes} we briefly review some relevant properties of NS crustal modes, and in Sec.~\ref{sec:QPOs}, we present the specific frequencies observed in FRB 20240114A. In Sec.~\ref{sec:oscillations}, we describe our NS models and the perturbation equations for the crustal torsional modes. Then, in Sec.~\ref{sec:NS_model}, the NS model for FRB 20240114A is estimated by identifying the observed frequencies with various $\ell$-th fundamental frequencies and 1st overtone of crustal torsional oscillations, considering the experimental constraint on the incompressibility $K_0$ of symmetric nuclear matter at saturation density. Simultaneously, through the resultant stellar model for FRB 20240114A, we derive a constraint on the density dependence of the nuclear symmetry energy at saturation density, the slope parameter $L$. Finally, we summarize our findings in this study and the conclusions in Sec.~\ref{sec:Conclusion}. Unless otherwise mentioned, we adopt geometric units with $c=G=1$, where $c$ and $G$ denote the speed of light and the gravitational constant, and use the metric signature $(-,+,+,+)$.


\section{Modeling the crustal modes of neutron stars}
\label{sec:modes}

The growing body of recent FRB observations provides an opportunity to further the study of NS physics and the cold dense matter equation of state (EOS) with new astrophysical data.
Because it is challenging to obtain information about nuclear properties at higher density regimes through terrestrial experiments, the EOS for NS matter is not yet fixed, but NS observations and inferences offer a path. For example, the discovery of massive NSs could exclude 
softer EOS models, with which the expected maximum mass does not reach the highest observed masses ~\citep{2010Natur.467.1081D,2021ApJ...915L..12F,2025ApJ...983L..20S,2026ApJ...996..101R}.  

Identifying FRB QPOs with NS crustal modes can add to the existing EOS constraints. Each perturbation which putatively causes a burst or FRB is thought to possess a spectrum with modes of varying amplitudes and phases depending on the details of the initial conditions of the perturbation that set the crust into motion. In most cases where FRB QPOs are observed (only a few percent of FRBs), a single dominant mode appears to exist or is observationally recoverable.

Since the oscillation frequencies of a NS strongly depend on the interior properties of the star, we can expect to constrain the NS properties by observing the frequencies (solving the inverse problem or, more appropriately, through forward modeling). This type of study is termed asteroseismology, in analogy to seismology on Earth and helioseismology on the Sun. Future observations of these oscillation frequencies, e.g., in the gravitational waves from a NS, will help constrain the NS mass, radius, and EOS \citep[e.g.,][]{PhysRevLett.77.4134,10.1046/j.1365-8711.1998.01840.x,10.1111/j.1365-2966.2005.08710.x, PhysRevD.83.024014,PhysRevD.88.044052}. This technique cannot yet be applied because gravitational wave detections of NS oscillation modes are not available, and magnetar QPOs therefore provide a practical alternative. In fact, by identifying the observed QPO frequencies with the crustal torsional oscillations, we can constrain the crust EOS and NS models \citep[e.g.,][]{10.1111/j.1365-2966.2011.19628.x,PhysRevLett.108.201101,SKS23}. However, most of the reported magnetar QPOs were observed in a few giant flares and intermediate bursts. In contrast, the accelerating rate of FRB detections could make FRBs competitive in providing additional NS constraints.

Whereas the oscillation frequency associated with the dynamical time of a NS is on the order of a few kHz, the QPO frequencies observed from magnetars range from tens of Hz to kHz. Therefore, theoretical identification of these frequencies, especially lower than $\sim 100$ Hz, is more difficult. The potential candidates may be crustal torsional modes or magnetic oscillations (or magneto-elastic oscillations). However, the magnetic (and also magneto-elastic) oscillations strongly depend on magnetic geometry and distribution of magnetic field strength inside the star~\citep{10.1093/mnras/sts721}, and also on how the magnetic fields are entangled~\citep[e.g.,][]{2016ApJ...823L...1L,2016ApJS..224....6L,2017APS..APR.Y4007B,2021MNRAS.504.5880B}, all of which are still quite uncertain, though such magnetic effects appear only if the field strength is significantly strong, as we discuss below.

In order to avoid such uncertainties from the magnetic fields, we consider a simple description of crustal torsional oscillations without any magnetic effects in this study. We note that this assumption is reasonable if the magnetic field is not extremely high, because the modification of the frequencies of the crustal torsional oscillations is only significant for magnetic fields stronger than $\sim \mathrm{few} \times 10^{15}$~G \citep{10.1093/mnras/sty445,2019PhRvD.100d3017D,SKS23}. 
That is, even for magnetar field strengths $\sim 10^{15}$~G, magnetic field corrections may be neglected, especially given other modeling uncertainties.

There is a further advantage to identifying the QPOs with the crustal torsional oscillations: because they are axial, i.e., they do not involve radial motion, they do not alter the shape of the star. Since torsional oscillations are excited only inside the NS crust, where the elasticity becomes non-zero, EOS uncertainties in the core have a minimum impact.\footnote{Uncertainties in the unknown core EOS translate into uncertainties in the inferred core mass and radius. In our approach, we integrate the TOV equations inward from the surface to the crust-core boundary, so the resulting NS (crust) models are specified by two parameters, 
$M$ and $R$. This contrasts with the standard outward integration from the center, where a fixed EOS yields a one-parameter family of stellar models.}
As the QPOs discovered in FRB 20240114A also lie within the same frequency range as the magnetar QPOs, we propose to identify/map the QPOs observed in FRB 20240114A with crustal torsional oscillations and derive constraints on stellar models and EOS parameters ~\citep{10.1093/mnras/sty1755,SKS23}. 
In \citet{SKS23}, the mass and radius of the NS associated with GRB 200415A were estimated with the identification of the reported higher QPO frequencies with the overtones of crustal torsional oscillations. In contrast to that work, here (owing to the existence of low-frequency QPOs as well as a high-frequency QPO candidate in FRB 20240114A), we estimate not only the mass and radius of the NS, but additionally constrain properties of symmetric nuclear matter at saturation density, by identifying both the low-order and higher QPO frequencies with the fundamental oscillations and overtones of crustal torsional oscillations.

\section{Quasi-periodic oscillations in FRB 20240114A}
\label{sec:QPOs}

FRB 20240114A is associated with a host galaxy at $z \approx 0.130287 \pm 0.000002$ ~\citep{10.1093/mnras/stae2013,2024ATel16613....1B,chen2025hostgalaxyhyperactiverepeating,2025A&A...695L..12B,2025ApJ...992L..35B}. 
The reciprocal of the (quasi-)periodic time intervals between bursts in a train observed in FRB 20240114A using FAST~\citep{zhou2025comprehensivesearchlongshort} provides an estimate of the observed lower quasiperiodic oscillation frequencies ($\lesssim 200$ Hz), $\nu_{\rm ob}$. The asymptotic frequency (for an observer far from the source) observed in the rest-frame of the source~\citep{2017LRR....20....7P},
 $\nu_0$, located at  redshift $z$, is estimated by
\begin{equation}
  \nu_0 = \nu_{\rm ob}(1+z), \label{eq:nu_ob}
\end{equation}
\citep{Wadiasingh_2020}, as listed in Table~\ref{tab:QPO1}.

\begin{table}
\caption{Lower-frequency QPOs: observed time interval between burst trains in FRB 20240114A, $\Delta t_{\rm ob}$, and its statistical significance, as reported in Figs.~4 and~D7 by ~\citet{zhou2025comprehensivesearchlongshort}. The observed frequency, $\nu_{\rm ob}$, estimated from $\Delta t_{\rm ob}$, and the frequency in the rest-frame of the object, $\nu_0$, using Eq.~(\ref{eq:nu_ob}), are also listed. The rightmost column is the corresponding value of $\ell$, obtained by identifying the frequencies $\nu_0$ with the $\ell$-th fundamental frequencies of crustal torsional oscillations (see \S~\ref{sec:NS_model1} for details).  } 
\label{tab:QPO1}
\begin {center}
\setlength{\tabcolsep}{5pt}
\begin{tabular}{ccccc}
\hline\hline
$\Delta t_{\rm ob}$ (ms)  & significance ($\sigma$) & $\nu_{\rm ob}$ (Hz) & $\nu_0$ (Hz) & $\ell$ \\ 
\hline
13.3    &  3.0 &  75.2  & 85.0  & 10 \\
30.6    &  3.2 &  32.7  & 36.9  &  4  \\
48.8    &  3.0 &  20.5  & 23.2  &  3  \\
62.6    &  3.9 &  16.0  & 18.1  &  2  \\
\hline
  5.5    &  1.8 & 181.8  & 205.6  & $> 13$ \\
11.5    & 1.6  & 87.0    &   98.3  & 11  \\
14.7    &  2.2 & 68.0    &   76.9  &  9  \\
21.4    &  1.9 & 46.7    &   52.8  &  6  \\
25.7    & 2.6  & 38.9    &   44.0  &  5  \\
29.5    & 2.8  & 33.9    &   38.3  &  4  \\
44.8    & 1.5  & 22.3    &   25.2  &  3  \\
\hline \hline
\end{tabular}
\end {center}
\end{table}

\citet{zhou2025comprehensivesearchlongshort} also report quasi-periodic oscillations in the few-hundred-Hz range for FRB 20240114A, presenting not only the central frequency, $\nu_{\rm ob}$, but also the full width at half maximum (FWHM$_{\rm ob}$). From these, we can estimate the corresponding values observed in the source rest-frame, $\nu_{0}$ and FWHM, via Eq.~(\ref{eq:nu_ob}). Moreover, using $\nu_{0}$ and FWHM, we can estimate the uncertainty in $\nu_0$ (from the lack of precise wave phase information) as
\begin{equation}
  \nu_{\rm min} < \nu_0 < \nu_{\rm max}, \label{eq:nu_range}
\end{equation}
where $\nu_{\rm min}$ and $\nu_{\rm max}$ are estimated from $\nu_0$ and the frequency widths, $\Delta \nu$,~\citep{Miller_2019} as
\begin{gather}
  \nu_{\rm min} = \nu_0 \left[1-\frac{1}{2}\left(\frac{\Delta \nu}{\nu_0}\right)^2\right],  \label{eq:nu_min}\\
  \nu_{\rm max} = \nu_0 \left[1+\frac{1}{2}\left(\frac{\Delta \nu}{\nu_0}\right)^2\right]. \label{eq:nu_max}
\end{gather}
In our case, $\Delta \nu$ is set to be half of the reported FWHM. The resulting values are listed in Table~\ref{tab:QPO2}. 
In the following Section, we separately consider the identification of the lower QPO frequencies listed in Table~\ref{tab:QPO1} and the higher QPO frequencies listed in Table~\ref{tab:QPO2} with fundamental frequencies and the 1st overtone of crustal torsional oscillations, respectively.

\begin{table*}
\caption{Higher-frequency QPOs: statistical significance, central frequency, $\nu_{\rm ob}$, the full width at half maximum (FWHM$_{\rm ob}$), and quality factors (Q), reported for FRB 20240114A in Table 2 of ~\citet{zhou2025comprehensivesearchlongshort} for a multi-QPO model. The QPO central frequency, $\nu_0$;  FWHM; and the minimum and maximum frequencies, $\nu_{\rm min}$ and $\nu_{\rm max}$, using Eqs.~(\ref{eq:nu_min}) and (\ref{eq:nu_max}), observed in the rest-frame of the object are also listed.
The rightmost column is the corresponding value of $(\ell,n)$, obtained by identifying the frequencies $\nu_0$ with the $\ell$-th frequencies of $n$-th overtone of the crustal torsional oscillations (see \S~\ref{sec:NS_model2} for details).}
\label{tab:QPO2}
\begin {center}
\setlength{\tabcolsep}{5pt}
\begin{tabular}{ccccccccc}
\hline\hline
significance($\sigma$) & $\nu_{\rm ob}$ (Hz)  & FWHM$_{\rm ob}$ (Hz) &  Q & $\nu_0$ (Hz) & FWHM (Hz) & $\nu_{\rm min}$ (Hz) & $\nu_{\rm max}$ (Hz) & $(\ell,n)$\\ 
\hline
 \multirow{2}{*}{3.4} & $266.85\pm 0.68$   &  $97.0\pm 1.3$   & $2.750\pm 0.037$  &  $301.7\pm 0.8$  & $106.7\pm 1.5$  & 295.8 & 307.6 & $(> 13,0)$\\
 & $502.1\pm 6.2$ &  $70.0\pm 14.0$  &    $7.10\pm 1.40$      &  $567.7\pm 7.0$  & $79.1\pm 15.8$   & 558.7 & 576.6 & $(\ell,1)$ \\
\hline
 \multirow{2}{*}{3.7} & $327.94\pm 0.93$   &  $117.4\pm 1.7$ &  $2.793\pm 0.041$  & $370.8\pm 1.05$   & $132.7\pm 1.9$  & 363.6 & 377.9 & $(> 13,0)$ \\
 & $579.8\pm 5.6$ &  $105.0\pm 10.0$ &   $5.50\pm 0.55$    & $655.5\pm 6.3$   & $118.7\pm 11.3$  & 645.9 & 665.0 & $(\ell,1)$  \\
\hline \hline
\end{tabular}
\end {center}
\end{table*}

\section{Crustal torsional oscillations}
\label{sec:oscillations}

To calculate the crustal torsional oscillations, we first prepare the background NS models. We consider a static, spherically symmetric spacetime, whose metric is given by
\begin{equation}
  ds^2 = -e^{2\Phi}dt^2 + e^{2\Lambda}dr^2 + r^2d\theta^2 + r^2\sin^2\theta d\phi^2, \label{eq:metric}
\end{equation}
where the metric functions, $\Phi$ and $\Lambda$, depend only on the radial coordinate, $r$, and $\Lambda$ is directly connected to the mass function, $m(r)$, as $e^{-2\Lambda}=1-2m/r$. We construct our stellar models  by integrating the TOV equations with an appropriate EOS for NS matter. 

The perturbation equation for crustal torsional oscillations is derived by linearizing the equation of motion and can be written as:
\begin{align}
  {\cal Y}''&+\left[\left(\frac{4}{r} + \Phi' - \Lambda'\right) + \frac{\mu'}{\mu}\right]{\cal Y}' \nonumber \\
      &+\left[\frac{\tilde{H}}{\mu}\omega^2e^{-2\Phi} -\frac{(\ell+2)(\ell-1)}{r^2}\right]e^{2\Lambda}{\cal Y}=0, \label{eq:perturbation}
\end{align}
where ${\cal Y}$ denotes the radial dependence of the angular displacement, and the prime denotes the derivative with respect to $r$, $\tilde{H}$ and $\mu$ are, respectively, the effective enthalpy density and the shear modulus defined below, and $\omega$ denotes the angular frequency of the $\ell$-th torsional oscillations \citep{10.1093/mnras/203.2.457,10.1093/mnrasl/sls006,10.1093/mnras/stt1152}. 
The boundary conditions to solve this equation are the vanishing of the traction force at the base and at the top of the crust, which reduce to ${\cal Y}'=0$ at these points \citep{10.1093/mnras/203.2.457,10.1111/j.1365-2966.2006.11304.x}. We note that one can determine not only the fundamental frequencies but also the overtones by solving the eigenvalue problem of Eq.~(\ref{eq:perturbation}), imposing the same boundary conditions.

For any EOS, the bulk energy per nucleon can be expanded in the vicinity of the saturation density, $n_0$,  of symmetric nuclear matter at zero temperature~\citep{annurev:/content/journals/10.1146/annurev-nucl-102711-095018} as a function of the baryon number density, $n_{\rm b}$, and neutron excess, $\alpha$, given by $\alpha=(n_n-n_p)/(n_n+n_p)$ with the neutron and proton number densities, $n_n$ and $n_p$, respectively:
\begin{equation}
  w=w_0 + \frac{K_0}{18n_0^2}\left(n_{\rm b}-n_0\right)^2 + \left[S_0 + \frac{L}{3n_0}\left(n_{\rm b}-n_0\right)\right]\alpha^2, \label{eq:w}
\end{equation}
where $w_0$ and $K_0$ are the saturation energy and incompressibility of symmetric nuclear matter at $n_{\rm b}=n_0$, while the coefficient in the term with $\alpha^2$ corresponds to the symmetry energy, $S(n_{\rm b})$. The symmetry energy is also expanded, as in Eq.~(\ref{eq:w}), where $S_0$ and $L$ are the symmetry energy and its density dependence at $n_{\rm b}=n_0$, i.e., $S_0=S(n_0)$ and $L=3n_0\left(dS/dn_{\rm b}|_{n_{\rm b}=n_0}\right)$. 

Among these five nuclear saturation parameters, $n_0$, $w_0$, and $S_0$ are relatively well constrained 
~\citep{RevModPhys.89.015007,Li:2019xxz}, while the remaining two parameters, $K_0$ and $L$, are experimentally poorly constrained. This is because $K_0$ and $L$ are the density \emph{derivatives} at $n_{\rm b}=n_0$, but the experimental data are concentrated around saturation density.
Nevertheless, experimental constraints on $K_0$ exist: $K_0=240\pm 20$ MeV \citep{2006EPJA...30...23S}. Meanwhile, we adopt a fiducial $L=60\pm 20$ MeV~\citep{PhysRevC.86.015803,refId0,Li:2019xxz}\footnote{Recent terrestrial experimental constraints on $L$ are not mutually consistent, e.g., PREX-II~\citep{PhysRevLett.126.172502,PhysRevLett.126.172503} or S$\pi$RIT~\citep{PhysRevLett.126.162701}.}. 

In order to systematically examine the torsional oscillations by changing the values of $K_0$ and $L$, we adopt the phenomenological EOS proposed by Oyamatsu and Iida~\citep{10.1143/PTP.109.631,PhysRevC.75.015801} (hereafter referred to as OI-EOSs) in this study. The OI-EOSs are constructed within the extended Thomas-Fermi theory in such a way that the bulk energy expression reduces to the expression given by Eq.~(\ref{eq:w}) in the limit of $n_{\rm b}\to n_0$ and $\alpha \to 0$, and that the values of $n_0$, $w_0$, and $S_0$ are optimized to reproduce experimental data for masses and charge radii of stable nuclei for given values of $K_0$ and $L$.

Crustal toroidal oscillations are restored due to the elasticity resulting from the crust’s crystalline structure, which is characterized by the shear modulus, $\mu$ (see eq. (\ref{eq:perturbation})). The shear modulus depends on the various phases of non-uniform nuclear structures. In the phase of body-centered cubic lattice of spherical nuclei, the effective shear modulus, $\mu_{\rm sp}$, has been derived as a function of nuclei number density, $n_i$, the charge number of nuclei, $Z$, and the radius of the Wigner-Seitz cell, $a$, as
\begin{equation}
  \mu_{\rm sp}=0.1194\frac{n_i(Ze)^2}{a} \label{eq:mu_sp}
\end{equation}
under the assumption that the nuclei are point particles, and taking the average over all possible wave vectors of displacements~\citep{PhysRevA.42.4867,1991ApJ...375..679S} and $1/n_i=4\pi a^3/3$. 

It is theoretically suggested that non-spherical atomic nuclei, the so-called pasta structures, exist at the base of the crust, where the shape of the nuclei changes from spherical to cylindrical, slab-like, cylindrical-hole, and spherical-hole before matter becomes uniform, as density increases. The shear modulus in the phase of cylindrical nuclei is estimated as
\begin{equation}
  \mu_{\rm cy}= \frac{2}{3}E_{\rm Coul}\times 10^{2.1(w_2-0.3)}, \label{eq:mu_cy}
\end{equation}
where $E_{\rm Coul}$ and $w_2$ denote the Coulomb energy per volume of a Wigner-Seitz cell and the volume fraction of cylindrical nuclei. The shear modulus in slab-like nuclei is zero, at least to linear order \citep{PETHICK19987}. That is, matter in the phase of slab-like nuclei behaves as a fluid in the linear perturbation regime. 
In this study, we consider crustal torsional oscillations excited in the phases of spherical and cylindrical nuclei only. Although there is a possibility for additional excitation of torsional oscillations in the phases of cylindrical-hole and spherical-hole nuclei, they would be independent of the torsional oscillations we focus on~\citep{10.1093/mnras/stz2385}.

Additionally, the fraction of dripped neutrons in the crust that behave as a superfluid, i.e. the ratio of superfluid neutrons to dripped neutrons $N_s/N_d$, also modifies the frequencies of crustal torsional oscillations, as shown in ~\citet{10.1093/mnrasl/sls006,10.1093/mnras/stt1152}. 
Namely, the case with $N_s/N_d=0$ is the situation in which all dripped neutrons comove with the nuclei, while the case with $N_s/N_d=1$ is the situation in which all dripped neutrons behave as a superfluid and do not contribute to the torsional oscillations. The information on $N_s/N_d$ is still poorly understood, but it is estimated in the phase of spherical nuclei by~\cite{PhysRevC.85.035801}, which depends on the density, and we adopt this estimation in this study. Meanwhile, since the value of $N_s/N_d$ inside the phase of cylindrical nuclei is poorly understood, we consider the extreme cases with $N_s/N_d=0$ and 1 for this phase.
The effective enthalpy density in Eq.~(\ref{eq:perturbation}), $\tilde{H}$, depends on the value of $N_s/N_d$ as $\tilde{H}=(1-N_s/A)H$, where $A$ denotes the baryon number inside a Wigner-Seitz cell and $H$ is the total enthalpy given by $H=\varepsilon + p$, where  $\varepsilon$ is the total energy density, and $p$ is the fluid pressure.


\section{Estimating a neutron star model for FRB 20240114A}
\label{sec:NS_model}

\subsection{Fundamental Torsional Oscillations}
\label{sec:NS_model1}

\begin{figure}[tbp]
\begin{center}
\includegraphics[width=0.49\textwidth]{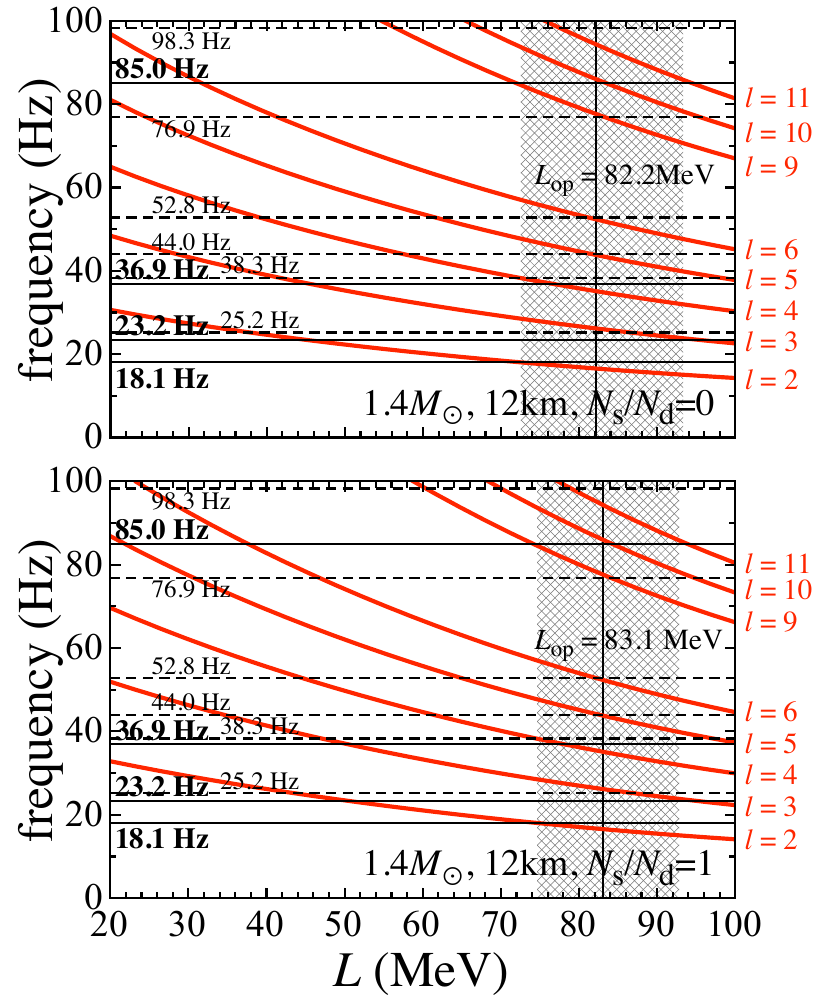} 
\end{center}
\caption{
The rest-frame frequencies, $\nu_0$, listed in Table~\ref{tab:QPO1} (except for 205.6 Hz), identified with the fundamental frequencies of crustal torsional oscillations with various values of $\ell$, for the NS model with $1.4_\odot$ and 12 km. The top and bottom panels correspond to the results with $N_s/N_d=0$ and 1, respectively. The horizontal solid lines with the QPO frequencies written in bold font are frequencies with statistical significance greater than or equal to 3$\sigma$, while the dashed lines with the QPO frequencies written in thin font are those with significance less than 3$\sigma$. The vertical lines denote the optimal value of $L$, (denoted by $L_{\rm op}$), to identify the QPO frequencies with the fundamental frequencies of the crustal torsional oscillations, while the shaded regions denote $L_{\rm op}\pm 1\sigma$.
}
\label{fig:M14R12}
\end{figure}

For a given NS model, with a specified mass, radius, $N_s/N_d$, and EOS parameters $K_0$ and $L$, the frequencies of the torsional oscillations are fully determined. 
However, it has been shown that the frequencies of the $\ell$-th fundamental torsional oscillations depend only weakly on $K_0$ and can therefore be characterized as a function of $L$ as
\begin{equation}
 {}_\ell t_0\, ({\rm Hz}) = c_\ell^{(0)} + c_\ell^{(1)}L_{100}^{1/2} + c_\ell^{(2)}L_{100} + c_\ell^{(3)}L_{100}^2, \label{eq:t0}
\end{equation}
where $L_{100}$ is defined as $L_{100}\equiv L/(100\ {\rm MeV})$ and $c_\ell^{(i)}$ for $i=0,1,2,3$ are fitting parameters that depend on the mass, radius, and $N_s/N_d$~\citep{PhysRevLett.108.201101,10.1093/mnrasl/sls006,10.1093/mnras/stt1152}\footnote{Here, we include an additional $L_{100}^{1/2}$ term in the function of $L$ for the fundamental frequencies (\ref{eq:t0}) for greater accuracy.}. 
larger than or equal to (smaller than) 3.0. 
In Fig.~\ref{fig:M14R12} we show as an example the correspondence between the QPO frequencies, $\nu_0$, listed in Table~\ref{tab:QPO1}, and the frequencies of fundamental torsional oscillations with specific values of $\ell$, calculated as a function of $L$ for the stellar model with $1.4M_\odot$ and 12 km. The solid (dashed) horizontal lines denote $\nu_0$ whose reported statistical significance is larger than or equal to (smaller than) 3$\sigma$. 
The upper and lower panels correspond to the results with $N_s/N_d=0$ and 1, respectively, and 
The coefficients in Eq.~(\ref{eq:t0}) are listed in Table~\ref{tab:c_ell}. 
Each QPO frequency in Table~\ref{tab:QPO1} can be consistently identified with one of the $\ell$-th crustal torsional modes, and the corresponding values of $\ell$ are also listed in Table~\ref{tab:QPO1}. The only exception is the 205.6 Hz QPO, which might be identified with a very high $\ell$ fundamental mode. 
We note that not all QPOs are observed at all times; only some are observed each time observations are made. That is, in this scenario, different torsional modes are excited at different times in the crust of the same stellar model. This could be caused by the fact that mode excitation depends on the initial conditions, i.e., how the crust is rung. For different FRBs even in the same source, the initial perturbations on the crust need not be identical.

\begin{table}
\caption{Coefficients in Eq.~(\ref{eq:t0}) for the fundamental modes with specific values of $\ell$ and for the NS model shown in Fig.~\ref{fig:M14R12}. } 
\label{tab:c_ell}
\begin {center}
\setlength{\tabcolsep}{5pt}
\begin{tabular}{cccccc}
\hline\hline
 $N_s/N_d$ & $\ell$ & $c_\ell^{(0)}$ & $c_\ell^{(1}$ & $c_\ell^{(2)}$ & $c_\ell^{(3)}$ \\ 
\hline
  0 &  2  & $44.07$ & $-26.94$ & $-7.979$ & $5.131$   \\
    &  3  & $69.68$ & $-42.58$ & $-12.65$ & $8.125$   \\
    &  4  & $93.47$ & $-57.08$ & $-17.01$ & $10.91$   \\
    &  5  & $116.6$ & $-71.15$ & $-21.24$ & $13.61$   \\
    &  6  & $139.3$ & $-84.94$ & $-25.48$ & $16.28$   \\
    &  9  & $206.4$ & $-125.3$ & $-38.45$ & $24.30$   \\
    & 10  & $228.6$ & $-138.6$ & $-42.81$ & $26.97$   \\
    & 11  & $250.7$ & $-151.7$ & $-47.33$ & $29.67$   \\
\hline
  1 &  2  & $44.86$ & $-17.49$ & $-22.87$ & $9.619$   \\
    &  3  & $70.89$ & $-27.52$ & $-36.30$ & $15.24$   \\
    &  4  & $95.08$ & $-36.88$ & $-48.71$ & $20.45$   \\
    &  5  & $118.6$ & $-46.01$ & $-60.71$ & $25.48$   \\
    &  6  & $141.7$ & $-54.75$ & $-72.79$ & $30.51$   \\
    &  9  & $209.8$ & $-80.35$ & $-108.7$ & $45.37$   \\
    & 10  & $232.3$ & $-88.53$ & $-120.8$ & $50.35$   \\
    & 11  & $254.7$ & $-96.62$ & $-133.0$ & $55.32$   \\
\hline \hline
\end{tabular}
\end {center}
\end{table}

\begin{figure*}[tbp]
\begin{center}
\includegraphics[scale=0.6]{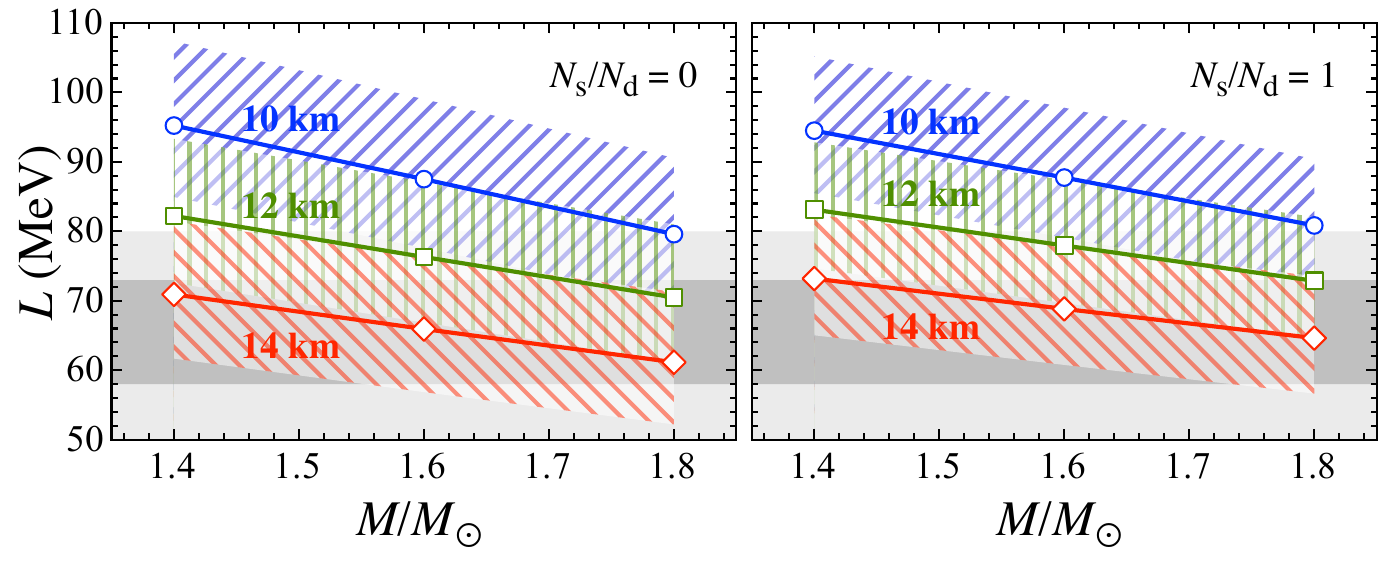} 
\end{center}
\caption{
The optimal ranges of $L_{\rm op} \pm 1\sigma$ to identify the observed QPO frequencies with the crustal torsional oscillations, shown for various NS models. The gray shaded regions denote the range of fiducial value of $L$, i.e., $L=60\pm 20$ MeV in light gray~\citep{PhysRevC.86.015803,refId0,Li:2019xxz}, and the range of $L$ constrained from the identification of magnetar QPOs observed in SGR 1806-20 with  crustal torsional oscillations, i.e., $L=58-73$ MeV in dark gray~\citep{10.1093/mnras/sty1755,universe10060231}.
}
\label{fig:LL_Ns}
\end{figure*}

Since the QPO frequencies reported in Table~\ref{tab:QPO1} are estimated from the time interval between the FRB burst trains, the lower limit in the uncertainty on $\Delta t_{\rm ob}$ is at the scale of the radio pulse widths, $\tau$. As that is not readily available, we (conservatively) adopt a uniform uncertainty in $\tau$. Then, $\nu_{\rm ob}$ is estimated as
\begin{equation}
  \Delta \nu_{\rm ob} \simeq \nu_{\rm ob}^2\tau. \label{eq:Df1}
\end{equation}
This choice gives more statistical weight to low-frequency QPOs, i.e., well-separated trains where the putative mode's period spacing between each $\ell$ is larger and identification is less ambiguous. We then propagate the uncertainty to the rest-frame of the object, $\Delta \nu_0$, as
\begin{equation}
  \Delta \nu_{0} = \Delta\nu_{\rm ob}(1+z)
  \simeq \nu_{\rm ob}^2(1+z)\tau. \label{eq:Df2}
\end{equation}
Here, we conservatively adopt $\tau\equiv3$~ms to estimate $\Delta \nu_{0}$.  This value is approximately the reported median of the effective pulse widths for FRB 20240114A with FAST~\citep{zhang2025investigatingfrb20240114afast}, and under our assumptions, is a systematic uncertainty associated with the times-of-arrival of a train of pulses within the radio burst.
The reported precision of the times-of-arrival of the pulses is considerably better than $\tau=3$~ms, however for broadened pulses (by choice of dispersion measure optimizing scheme, for instance), the intervals between observed peaks of pulses may not precisely map to the underlying physical mode. A future analysis could improve upon these assumptions by using morphological information of bursts in an instrument and frequency-specific way\footnote{For instance, optimizing for time varying epoch-to-epoch dispersion measure variations (for a repeating FRB) in a bespoke manner.}.  

The optimal value of $L$ to identify the (low-order) QPO frequencies, using the theoretical value $\nu_{\rm fit}$ (given by Eq.~\ref{eq:t0}), is determined so that the value of ${\cal F}$ defined as 
\begin{equation}
  {\cal F}(L)\equiv\sum_{i } \frac{(\nu_{{\rm 0},i}-\nu_{{\rm fit},i})^2}{\Delta \nu_{{\rm 0},i}^2} 
\end{equation}
is minimized, assuming the correspondence between each pair $(\nu_{0,i},\ell)$, obtained by identifying the $i$-th QPO frequency with one of the $\ell$-th torsional oscillation frequencies given by $\nu_{{\rm fit},i}$,  in such a way that the lowest candidate frequency, i.e., 18.1 Hz, is anchored to the $\ell=2$ fundamental torsional mode, as listed in Table~\ref{tab:QPO1}. Accordingly, the optimal value of $L$ for the stellar model with $1.4M_\odot$ and $12$ km is obtained as 82.2 MeV for $N_s/N_d=0$ and 83.1 MeV for $N_s/N_d=1$, as shown in Fig.~\ref{fig:M14R12}, where we also show the $1\sigma$ uncertainty in $L$ with the shaded region. Since the 10 QPOs listed in Table~\ref{tab:QPO1} (except for 205.6 Hz) evaluated in the rest frame are identified with the various fundamental torsional oscillations using one free parameter ($L$), the number of degrees of freedom is 9. So, the $1\sigma$ uncertainty in $L$ from the $\chi^2$ statistic requires ${\cal F}(L) -{\cal F}(L_{\rm op}) \approx10.42$, with $L_{\rm op}$ being the optimal value of $L$. If one restricts the QPOs listed in Table~\ref{tab:QPO1} to those whose reported significance is larger than 3$\sigma$, the optimal value of $L$ is almost unchanged (less than $0.1\%$), and the maximum (minimum) value of $L$ within the $\pm 1\sigma$ range becomes at most around $4\%$ smaller (larger). For $\tau =3$ ms, the minimum and maximum values of $L$ within $\pm 1\sigma$ are $72.5$ and $93.3$ MeV ($74.6$ and $92.8$ MeV) for $N_s/N_d=0$ ($N_s/N_d=1$) via Eq.~(\ref{eq:Df2}). With $\tau =1.5$ ms, these values are $77.2$ and $87.5$ MeV ($78.7$ and $87.8$ MeV) for $N_s/N_d=0$ ($N_s/N_d=1$). For $\tau =6$ ms, they are $64.0$ and $106.5$ MeV ($67.1$ and $104.5$ MeV) for $N_s/N_d=0$ ($N_s/N_d=1$). See also Appendix~\ref{sec:appendix_1} for a more extended analysis of the dependence of our results on the choice of $\tau$. 

The optimal range for $L$ varies with the stellar mass and radius because the fundamental frequencies depend on the stellar model. 
In Fig.~\ref{fig:LL_Ns}, we show $L_{\rm op} \pm 1\sigma$ for several stellar models, together with the fiducial value of $L$ obtained from experiments, i.e., $L=60\pm 20$ MeV~\citep{PhysRevC.86.015803,refId0,Li:2019xxz}, and the values constrained by the identification of the magnetar QPO frequencies observed in SGR 1806--20 as crustal torsional oscillations, i.e., $L=58-73$ MeV~\citep{10.1093/mnras/sty1755,universe10060231} with shaded regions. From Fig.~\ref{fig:LL_Ns}, we find that stellar models with a radius of 10 km and a mass less than $\sim 1.6M_\odot$ are not consistent with the fiducial value of $L$, irrespective of the value of $N_s/N_d$ in the phase of the cylindrical nuclei. In addition, we find that larger values of $L$ are favored for models with smaller radius, for fixed stellar mass.

\subsection{Overtones}
\label{sec:NS_model2}

\begin{figure}[tbp]
\begin{center}
\includegraphics[width=0.45\textwidth]{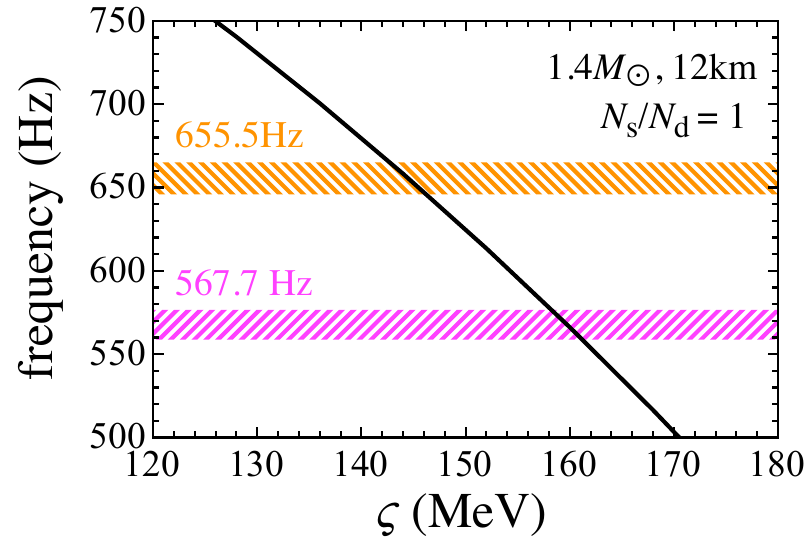} 
\end{center}
\caption{
The $\sim 600$ Hz candidate QPO frequencies observed in FRB 20240114A, compared with the 1st overtone excited in a NS model with $1.4M_\odot$, 12 km, and $N_s/N_d=1$.
}
\label{fig:1st-M14}
\end{figure}

Now we turn to the higher-frequency QPO candidates shown in Table~\ref{tab:QPO2}. These QPOs may be identified with the first overtone of the crustal torsional oscillations, which have been shown to be almost independent of $\ell$, but depend on $K_0$ and $L$. This is because the overtone frequencies depend on the crust thickness~\citep{1980ApJ...238..740H}, and the crust thickness depends on $K_0$, $L$, and $M/R$~\citep{10.1093/mnras/stx1510}. The overtone frequencies are well characterized by $\varsigma$ defined as
\begin{equation}
  \varsigma \equiv \left(K_0^4L^5\right)^{1/9}. \label{eq:sigma}
\end{equation}
In practice, as shown in \citet{10.1093/mnras/sty1755,SKS23}, the overtone frequencies of crustal torsional oscillations are expressed as a function of $\varsigma$ as
\begin{equation}
  {}_\ell t_n\, {\rm (Hz)} = d^{(0)}_{\ell n} + d^{(1)}_{\ell n}\varsigma_{100} + d^{(2)}_{\ell n}\varsigma_{100}^2, \label{eq:tn}
\end{equation}
where $\varsigma_{100}$ is defined as $\varsigma/(100\ {\rm MeV})$, and $d^{(i)}_{\ell n}$ for $i=0,1,2$ are fitting parameters depending on the stellar mass, radius, and $N_s/N_d$, but almost completely independent of $\ell$. Thus, once an identification of a high-frequency QPO is made with the overtone, a constraint on $\varsigma$ is obtained. This constraint is independent of the constraint on $L$ obtained through identification of the fundamental oscillations, shown in Fig.~\ref{fig:LL_Ns}.

\begin{table*}
\caption{The allowed maximum and minimum masses for the NS models with 14 and 12 km and with $N_s/N_d=0$ and 1, constrained by identifying the $567.7$ Hz or $655.5$ Hz QPOs with the 1st overtone of crustal oscillations and by identifying the lower QPOs with the fundamental torsional oscillations, together with the terrestrial constraint on $K_0$. In addition, the value of $L_{\rm max}$ ($L_{\rm min}$) corresponding to the NS model with the allowed maximum mass (minimum mass) is also listed. } 
\label{tab:NS_mass}
\begin {center}
\setlength{\tabcolsep}{5pt}
\begin{tabular}{ccccccc}
\hline\hline
QPO (Hz)  & $N_s/N_d$ & $R$ (km) & $M_{\rm max}/M_\odot$ & $M_{\rm min}/M_\odot$ & $L_{\rm max}$ (MeV) & $L_{\rm min}$ (MeV) \\ 
\hline
567.7   &  0 &  14  & 1.55  & 1.29 & 77.5 & 64.3 \\
    &    &  12  & 1.29  &  1.04 & 96.8 & 82.9 \\
    &  1 &  14  & 1.54  &  1.27 & 79.3 & 67.7 \\
    &    &  12  & 1.27  &  1.00 & 96.4 & 84.7 \\
\hline
655.5   &  0 & 14  & 1.76  & 1.49 & 72.1 & 59.5 \\
    &    & 12   &   1.46  & 1.19 & 91.5 & 78.4 \\
    &  1 & 14   &   1.75  & 1.48 & 74.5 & 63.3 \\
    &    & 12   &   1.44  & 1.17 & 91.7 & 80.5 \\
\hline \hline
\end{tabular}
\end {center}
\end{table*}

\begin{figure*}[tbp]
\begin{center}
\includegraphics[scale=0.52]{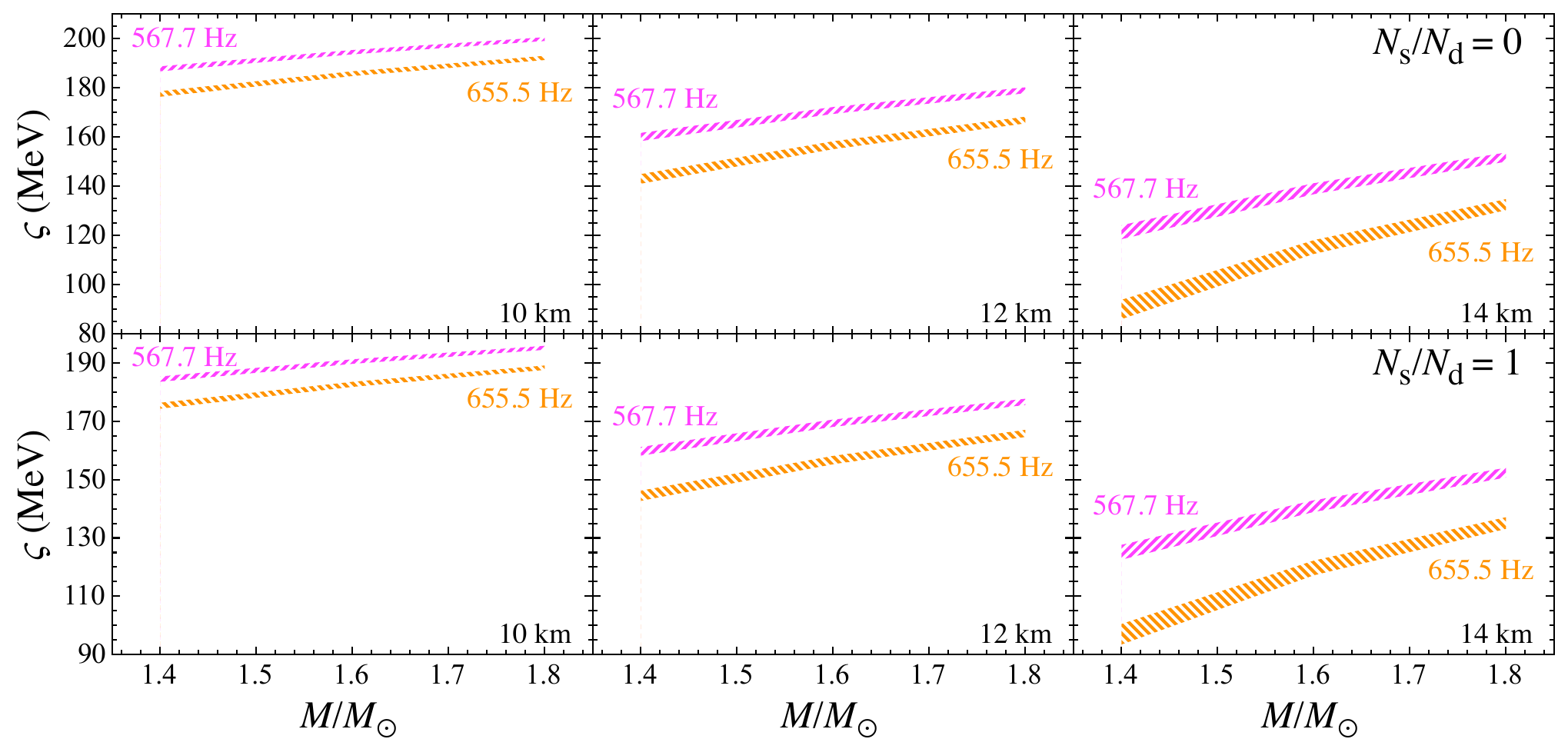} 
\end{center}
\caption{
Constraints on the parameter $\varsigma$, obtained by identifying the QPO frequencies of $\sim 600$ Hz with the 1st overtone of crustal torsional oscillations. The left, middle, and right panels correspond to stellar models whose radii are 10, 12, and 14 km, respectively, while the top and bottom panels correspond to results with $N_s/N_d=0$ and 1. In each panel, the lower and upper bounds come from the upper and lower bounds of the uncertainties in the QPO frequency (see Fig.~\ref{fig:1st-M14}). 
}
\label{fig:ss_RNs10}
\end{figure*}

Next, we attempt to identify the frequencies listed in Table~\ref{tab:QPO2} except for the 301.7 and 370.8 Hz QPOs, as the 1st overtone of the crustal torsional oscillations.\footnote{The 301.7 and 370.8 Hz QPOs in Table~\ref{tab:QPO2} as well as the 205.6 Hz QPO in Table~\ref{tab:QPO1} could be considered as a result of the fundamental torsional oscillations with higher values of $\ell$. However, given their comparatively low statistical significance, these identifications should be treated as tentative and would have limited weight in the overall mode assignment.} 
In Fig.~\ref{fig:1st-M14}, we show as an example the comparison of the QPO frequencies with the 1st overtone as a function of $\varsigma$ for the stellar model with $1.4M_\odot$, 12 km, and $N_s/N_d=1$, and the coefficients in Eq.~(\ref{eq:tn}) are $d_{\ell n}^{(0)}=1042$, $d_{\ell n}^{(1)}=11.48$, and $d_{\ell n}^{(2)}=-193.0$.
Through the identification of the $\sim 600$ Hz QPO frequencies with the 1st overtone, we can obtain a constraint on $\varsigma$ for each stellar model. The resulting constraints on $\varsigma$ are shown in Fig.~\ref{fig:ss_RNs10} for varying  stellar mass and radius with $N_s/N_d=0$ (top panel) or 1 (bottom panel). 

We note that both 1st overtone candidates at 567.7 Hz and 655.5 Hz cannot be identified simultaneously in our non-magnetized crust formalism with different crustal torsional modes  \citep[if the reported uncertainties are not underestimated in][]{zhou2025comprehensivesearchlongshort}. This is because the model overtones are almost independent of the azimuthal quantum number, $\ell$, and the frequency difference between overtones is much larger than the difference between 567.7 Hz and 655.5 Hz (e.g., the frequency of the 2nd overtone is approximately 1 kHz). Since the frequency depends on the chosen (global) background model, one could simultaneously account for both frequencies if, for example, the stellar mass and/or magnetic field structures were to change. However, these two candidate QPOs are observed less than one day apart, and it is difficult to envision such significant and quick changes of the global NS field/crust structure ($\Delta M \sim \mathcal{O}(0.1 M_{\odot})$ and/or $\Delta B \sim \mathcal{O}(10^{15}\rm{G})$). 

Accordingly, we assume that only one of the two higher-frequency QPO candidates originates from the 1st overtone, whereas the other QPO may arise from a different NS mode, such as a shear or gravity mode, or a high-$\ell$ fundamental crustal torsional mode, imprinted on the signal. Therefore, we analyze each candidate overtone separately. 

\begin{figure*}[tbp]
\begin{center}
\includegraphics[scale=0.52]{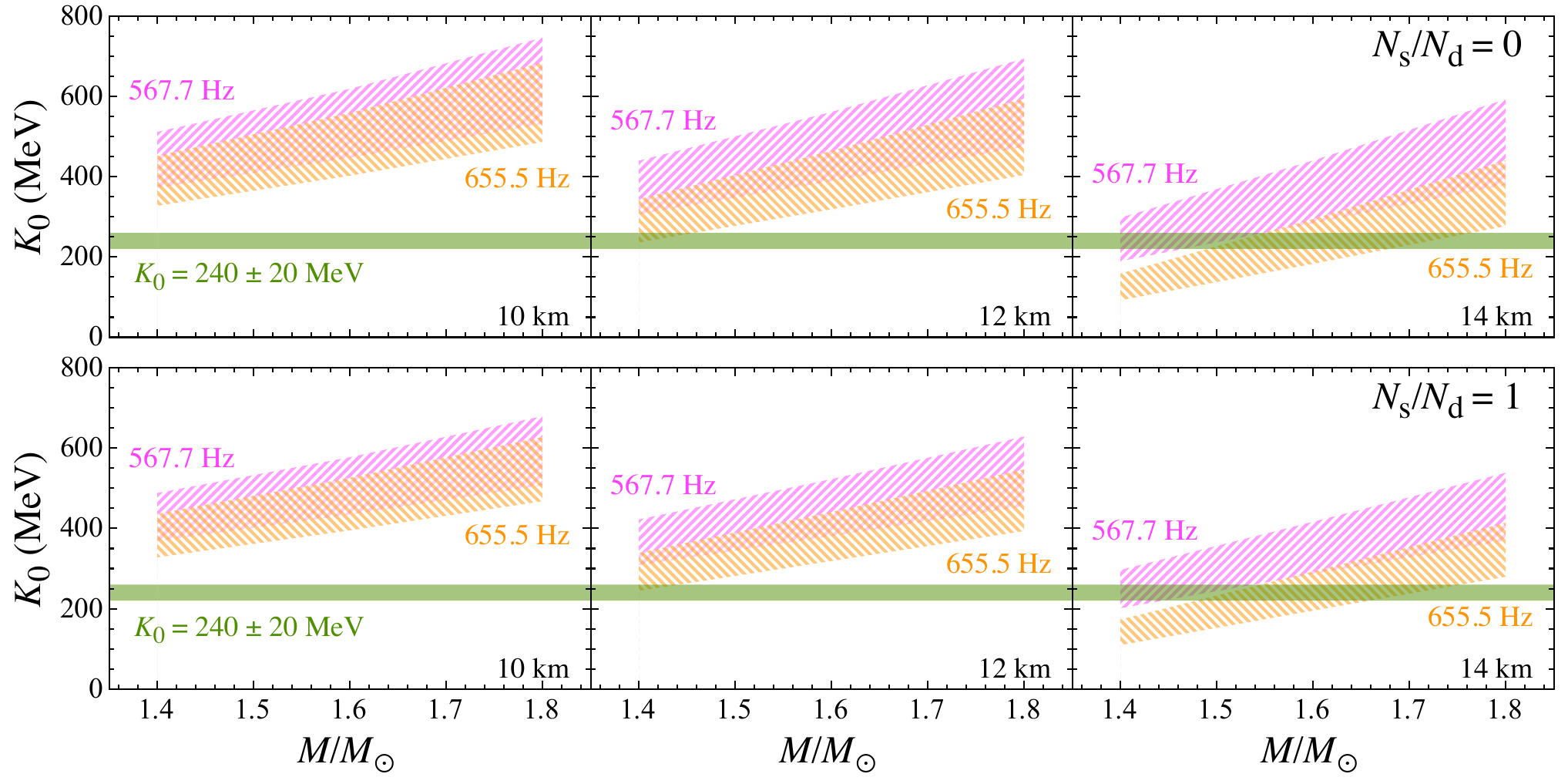} 
\end{center}
\caption{
Constraint on $K_0$ obtained from the combination of the constraints on $L$ shown in Fig.~\ref{fig:LL_Ns} and $\varsigma$ shown in Fig~\ref{fig:ss_RNs10} by identifying the QPO frequencies with the crustal torsional oscillations. For reference, the fiducial value of $K_0$ obtained from the terrestrial experiments, i.e., $K_0=240\pm20$ MeV \citep{2006EPJA...30...23S}, is also shown.
}
\label{fig:K0_Ns}
\end{figure*}

The combination of the constraint on $L$ from the identification of the lower QPO frequencies with the fundamental torsional oscillations, shown in Fig.~\ref{fig:LL_Ns}, and the constraint on $\varsigma$ from the identification of higher QPO frequency with the 1st overtone, shown in Fig.~\ref{fig:ss_RNs10}, gives us a constraint on $K_0$ via
\begin{equation}
  K_0=(\varsigma^9/L^5)^{1/4}. \label{eq:K0}
\end{equation}
In Fig.~\ref{fig:K0_Ns} we present our constraint on $K_0$, together with the fiducial value of $K_0$ from experiments, $K_0=240\pm20$ MeV \citep{2006EPJA...30...23S}. The upper (lower) bound of $K_0$ in Fig.~\ref{fig:K0_Ns} comes from the upper (lower) bound of $\varsigma$ in Fig.~\ref{fig:ss_RNs10} and the lower (upper) bound of $L$ in Fig.~\ref{fig:LL_Ns} for each stellar model. The resulting ranges for $K_0$, obtained by the identification of the QPOs with crustal torsional oscillations, can be compared with the values from experiments, and used to constrain the NS models.

In practice, focusing on the stellar models with 14 and 12 km, we obtain the maximum and minimum masses to identify the 567.7 Hz or 655.5 Hz QPO with the 1st overtone. The resulting masses for $N_s/N_d=0$ or 1 are listed in Table~\ref{tab:NS_mass}, where the masses smaller than $1.4M_\odot$ are estimated by extrapolating the regions shown in Fig.~\ref{fig:K0_Ns}. For a fixed stellar radius, the maximum and minimum masses constrained with $N_s/N_d=0$ become a little larger than those with $N_s/N_d=1$ for both cases with 567.7 and 655.5 Hz QPOs. Considering the uncertainties in $N_s/N_d$, the maximum mass constrained with $N_s/N_d=0$ and the minimum mass with $N_s/N_d=1$ gives us the allowed region of mass for the NS models with a fixed radius. 

\subsection{Mass-Radius Constraints}

\begin{figure*}[tbp]
\begin{center}
\includegraphics[width=0.99\textwidth]{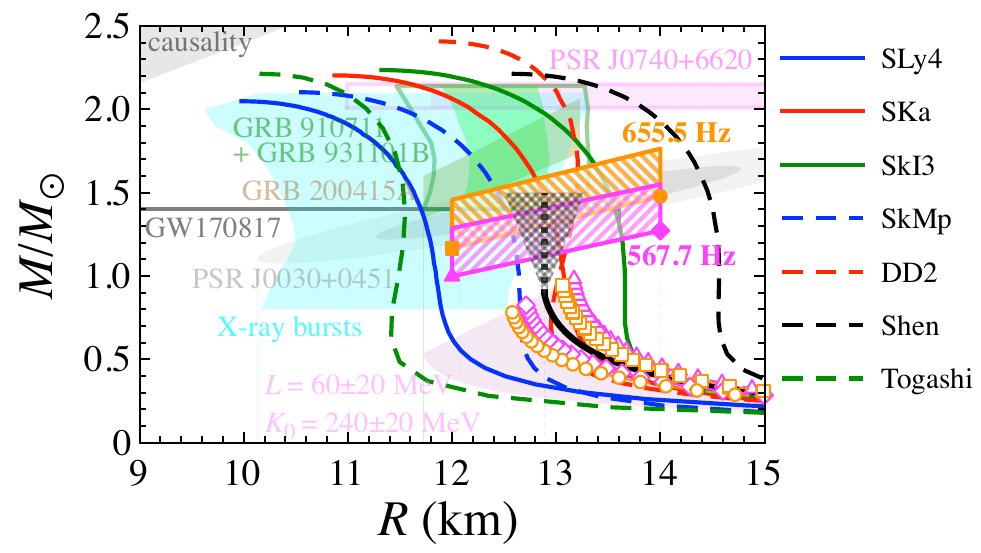} 
\end{center}
\caption{
The mass and radius of FRB 20240114A, constrained from the identification of the QPO frequencies with the crustal torsional oscillations, together with the constraint on the nuclear parameter obtained from the terrestrial experiments, are shown with the shaded regions with hashed lines, where the shaded region with hashed lines from top left to bottom right (with hashed lines from top right to bottom left) is the allowed region with the identification of the 655.5 Hz (567.7 Hz) with the 1st overtone. The sequences with open circles, open squares, open diamonds, and open triangles denote the NS mass and radius estimated from the low-mass formula~\citep{10.1093/ptep/ptu052}, using the same values of $L$ and $K_0$ for the stellar models marked with the filled circle, filled square, filled diamond, and filled triangle, respectively. For reference, we also show the other astronomical constraints on the NS mass and radius for PSR J0030+0451, PSR J0740+6620, GW170817, GRB 200415A, a combination of GRB 910711 and GRB 931101B and x-ray bursts; the theoretical constraint from causality; and the expected mass and radius from the fiducial values of $L$ and $K_0$ constrained from the experiments (see text for details).
In addition, the stellar models shown with a black solid line are the low-mass NS sequence, assuming $L=74$ MeV, which is a median value for the stellar models shown with open diamonds and squares, while the black dotted line is just an extension of the black solid line by fixing the radius ($R=12.9$ km). 
}
\label{fig:MR}
\end{figure*}

Following the procedure outlined above, we obtain the constraint on the mass and radius as viable NS models for FRB 20240114A, shown in Fig.~\ref{fig:MR}. The shaded region with hashed lines from top left to bottom right (with hashed lines from top right to bottom left) is the allowed range obtained with the identification of the 655.5 Hz (567.7 Hz) with the 1st overtone. For reference, we also plot other constraints obtained from astronomical observations, i.e., the NS mass and radius for PSR J0030+0451~\citep{2019ApJ...887L..24M} and PSR J0740+6620~\citep{2024ApJ...974..295D} constrained via x-ray observations with NICER (see also \cite{2019ApJ...887L..21R,2021ApJ...918L..27R,2019ApJ...887L..24M}); the tidal deformability constrained by GW170817~\citep{PhysRevLett.119.161101}, which leads to the constraint on the $1.4M_\odot$ NS radius being $\lesssim 13.6$ km~\citep{PhysRevLett.120.172703}; the estimated mass and radius by identifying the magnetar higher QPOs from GRB 200415A \citep{SKS23}; the $1\sigma$ and $2\sigma$ radius constraints, for neutron star masses between $1.4$ and $2.1M_\odot$, obtained by identifying the kHz QPOs from short GRBs 910711 and 931101B \citep{2023Natur.613..253C} with the quasi-radial and quadrupolar oscillations of a binary neutron star merger remnant~\citep{Guedes_2025}, and observations of x-ray bursts~\citep[for example,][]{2013ApJ...765L...5S}. The top-left shaded region is excluded by causality~\citep{annurev:/content/journals/10.1146/annurev-nucl-102711-095018}, while the bottom-right shaded region denotes the NS mass and radius estimated from the fiducial values of the nuclear saturation parameters, i.e., $K_0=240\pm 20$ MeV and $L=60\pm 20$ MeV, using the mass formula for a low-mass NS discussed in the next paragraph \citep{10.1093/ptep/ptu052}. Additionally, the mass and radius for NS models constructed with several realistic EOS are shown with solid and dashed lines.

In Fig.~\ref{fig:MR}, the filled circle and filled square (filled diamond and filled triangle) denote the NS models with the lowest mass with 14 km and 12 km radius, respectively, obtained by identifying the 655.5 Hz (567.7 Hz) QPO with the 1st overtone of crustal torsional oscillations. Each of these points should be connected, through a series of equilibrium models, to the corresponding limit provided by the low-density EOS, which can be obtained as follows. \citet{10.1093/ptep/ptu052} suggest that the mass and gravitational redshift for low-mass NS models, whose central density is less than twice the nuclear saturation density, are well expressed as a function of the central density and a combination of the nuclear saturation parameters  defined as
\begin{equation}
  \eta \equiv (K_0L^2)^{1/3}. \label{eq:eta}
\end{equation}
Using the relations provided by \citet{10.1093/ptep/ptu052}, we can estimate the mass and radius for low-mass NS models. We find that the mass for a NS model with fixed $L$ (central density) increases as the central density ($L$) increases, which leads to
$L$ increasing with radius. This trend is opposite to the result obtained from identifying the lower QPOs with the fundamental torsional oscillations. In fact, the NS mass and radius thus estimated, using the same value of $L$ and $K_0=220$ MeV for the stellar models denoted with the filled circle, filled square, filled diamond, and filled triangle (see $L_{\rm min}$ in Table~\ref{tab:NS_mass}), are shown with the open circles, open squares, open diamonds, and open triangles in Fig.~\ref{fig:MR}. So, for example, in a sequence of models determined by the same EOS, the stellar model denoted by the filled circle should be connected to the leftmost open circle. Therefore, we may be able to estimate the NS mass with a radius smaller than 12 km or larger than 14 km by identifying the higher QPO with the 1st overtone, but such a stellar model would be more difficult to connect to the lower-mass NS sequence. 

Considering the necessary interpolation to the lower-mass sequence, the radius of the NS model for FRB 20240114A is $\sim 13$ km. A possible sequence of low-mass NSs with $L=74$ MeV (the median value for the stellar models shown with open diamonds and squares), is shown in Fig.~\ref{fig:MR} with the solid black line, and it is naturally connected to the allowed region that we constrained in this study. As an illustrative example, we extend this sequence with the dotted line, which has a fixed stellar radius ($R=12.9$ km). In more general cases, the dotted line curves left or right, depending on the EOS. A very sharp turn left or right is not expected in realistic EOS, and thus the data limits the viable self-consistent range of the radius to be around $\sim 13$~km.

We note that the estimation of mass and radius for GRB 200415A is obtained by identifying only the higher QPOs with the overtones \citep{SKS23}. In contrast, the detection of \emph{both} the low and high frequency QPOs in FRB 2024114A enables us to estimate the mass and radius, and additionally set a constraint on $L$. The imprinting physics that connects the central engine (and neutron star crustal oscillations) to observable electromagnetic radiation is still uncertain and is likely different for radio (FRBs) and high-energy (GRBs) radiation. However, it is quite interesting that both observations in GRB 200415A and FRB 2024114A are explained within the same framework, which may suggest that both objects are magnetars, and the observed QPOs are associated with crustal torsional oscillations.

\begin{figure}[tbp]
\begin{center}
\includegraphics[scale=0.62]{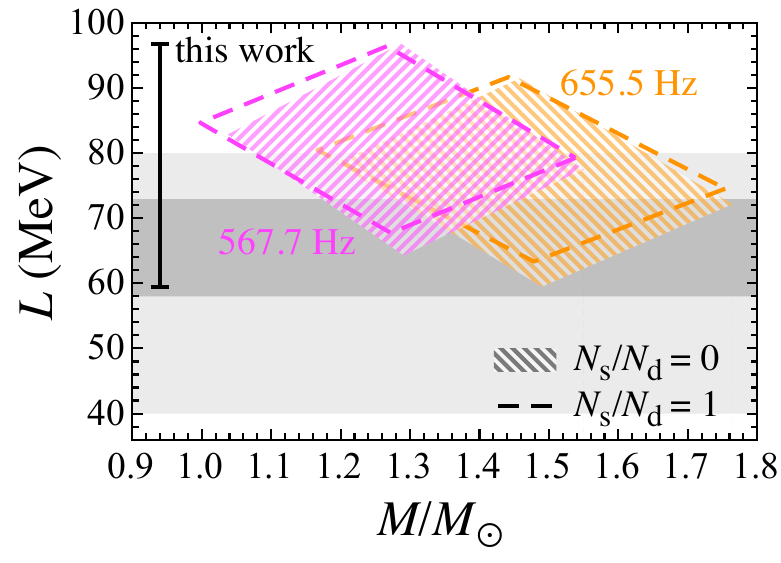} 
\end{center}
\caption{
Constraint on $L$ obtained by identifying the QPO frequencies with crustal torsional oscillations, together with the experimental constraint on $K_0$. The shaded region with hashed lines and the region enclosed by the dashed line denote the constraints on $L$, for  stellar models with $N_s/N_d=0$ and 1, respectively. From our analysis, we find $L=59.5-96.8$ MeV. The fiducial value of $L$ from  experiments, $L=60\pm 20$ MeV~\citep{PhysRevC.86.015803,refId0,Li:2019xxz}, and the constraint obtained from the analysis of magnetar QPOs, $L=58-73$~MeV~\citep{10.1093/mnras/sty1755,universe10060231}, are also shown with the light and dark gray shaded regions, respectively.
}
\label{fig:ML}
\end{figure}

Our results also provide constraints on the value of $L$. Whereas in Fig.~\ref{fig:MR} we focus on the mass constraints for stellar models with a given radius, in Fig.~\ref{fig:ML} we show the corresponding constraints on $L$, given in  Table~\ref{tab:NS_mass}. 
%
We also show the fiducial value of $L$, i.e., $L=60 \pm 20$ MeV (from experiments), as well as the constraint obtained from the identification of magnetar QPOs by ~\citet{10.1093/mnras/sty1755,universe10060231}, i.e., $58-73$ MeV. The allowed values of $L$ obtained from this study are in the range of $L=59.5-96.8$ MeV, which is broadly consistent with the fiducial value of $L$ and also with the constraint from the magnetar QPOs. Since this allowed range of $L$ depends on the stellar mass (and radius), it would become more restrictive if constraints on the mass and/or radius of the source (FRB 20240114A) were obtained through other observations.

\section{Conclusion}
\label{sec:Conclusion}

FRBs are energetic phenomena that we observe mostly from extragalactic distances. The QPOs recently discovered in FRB 20240114A motivate our systematic study of the crustal torsional oscillations that can be excited in a NS. The lower-frequency QPOs can be identified with the fundamental oscillations with various azimuthal quantum numbers, enabling us to constrain the density dependence of the nuclear symmetry energy slope parameter, $L$. On the other hand, the higher-frequency QPO is considered as a result of the 1st overtone, and its identification leads to another constraint on the combination of the nuclear saturation parameters, $\varsigma=(K_0^4L^5)^{1/9}$. This constraint on $\varsigma$ together with the constraint on $L$, obtained from the identification of the lower QPOs, produces a constraint on the incompressibility coefficient $K_0$, depending on the stellar model. However, $K_0$ is constrained by various terrestrial experiments, which prove to be more stringent than the constraint obtained in this study. That is, the experimental constraint on $K_0$ can select stellar models consistent with this interpretation for the QPOs. 

We caution, however,  that if the observed QPOs are related to magneto-elastic oscillations (i.e., not purely crustal) due to the presence of a magnetic field higher than a few $\times 10^{15}$~G, then a mapping to non-magnetic crust modes can become systematically biased. The inferred $L$ and the inferred $M$-$R$ region could shift or widen substantially. 
We note that the remarkably consistent results we have obtained from a mapping to non-magnetic crustal modes suggest either global fields lower than few$\times10^{15}$~G or perhaps stronger core-confined fields with only a limited region of active crust field evolution \citep[e.g.,][]{2024MNRAS.535.2449L,Lander_2026} powering activity. Indeed, in the dynamical Meissner model of \cite{Lander_2026}, only neutron stars with internal fields $\lesssim 5-7\times10^{14}$~G will likely have ``standard magnetar" crust evolution to power activity, which is well within the domain of applicability of non-magnetic crustal mode models. Meanwhile, studies of nearby extragalactic magnetar giant flares suggest birth fields at formation (before decay) of order $4\times 10^{14} -2  \times 10^{15}$~G \citep{2025ApJ...980..211B}, still possibly in the non-magnetic crust mode domain. 

Finally, even if some individual radio QPOs are of marginal statistical significance, in aggregate they may be significant.  This can only be tested by future large samples and families of FRB QPOs, or strong harmonic clustering in FRB short waiting time distributions, over a broad range of cosmological redshifts. This may be feasible with DSA~\citep{2019BAAS...51g.255H} and CHORD~\citep{2019clrp.2020...28V}. If QPO-like periodicities are confirmed in multiple repeaters, the field is likely to shift from single-source inversions to population-based seismology, constraining EOS parameters (assuming there is a single EOS that describes cold dense matter) statistically across sources.In the end, why particular individual modes are preferentially excited, imprinted \citep[e.g.,][]{2025ApJ...995L..57B} or observationally recoverable over others is an open question, which may be source, geometry or crust perturbation specific even without instrumentation selection biases. We note that the hard X-ray giant flare tail of SGR 1806-20 also exhibits clustering of transient frequencies with observability biased by rotational phase \citep{Miller_2019}.

\begin{acknowledgments}
This work is supported in part by the Japan Society for the Promotion of Science (JSPS) KAKENHI Grant Numbers 
JP23K20848       
and JP24KF0090.  
The material is based upon work supported by NASA under award numbers 80GSFC21M0002 and 80GSFC24M0006. We thank Paz Beniamini and Cole Miller for interesting discussions and Victor Guedes for sharing the data from the analysis of the BATSE QPOs.  This work has made use of the NASA Astrophysics Data System. 
\end{acknowledgments}


\appendix
\section{Dependence of the final results on the selection of $\tau$}   
\label{sec:appendix_1}

As in Table~\ref{tab:NS_mass}, Tables~\ref{tab:NS_mass_tau15} and \ref{tab:NS_mass_tau60} list the allowed ranges of maximum and minimum masses for the NS models with $14$ and 12 km and with $N_s/N_d=0$ and 1, but with $\tau = 1.5$~ms and $\tau = 6$~ms, respectively. These results are obtained by identifying 567.7 or 655.5 Hz QPOs with the 1st overtone of crustal torsional oscillations and by identifying the lower QPOs with the fundamental oscillations with various $\ell$, incorporating experimental constraints on $K_0$. The corresponding bounds on $L$, $L_{\rm max}$ and $L_{\rm min}$, are also listed. In both tables, the bracketed numbers following each entry denote the fractional deviation from the corresponding value for $\tau=3.0$~ms (shown in Table~\ref{tab:NS_mass}). The allowed maximum and minimum masses vary by a few $\%$ with $\tau =1.5$ ms, see Table~\ref{tab:NS_mass_tau15}; the allowed maximum (minimum) masses vary by up to $18\%$ (a few $\%$) with $\tau=6.0$ ms, see Table~\ref{tab:NS_mass_tau60}.

As a result, doubling or halving the uncertainty $\tau$ results in $\lesssim 10\%$ changes in most cases. The NS mass ranges obtained from the identification of the 567.7 or 655.5 Hz QPOs with the 1st overtone become, respectively,  $1.02-1.50M_\odot$ or $1.19-1.71M_\odot$ with $\tau=1.5$ ms and $0.96-1.66M_\odot$ or $1.12-2.08M_\odot$ with $\tau=6.0$ ms. Meanwhile, the constraint on $L$ from the optimal stellar models for explaining the QPO observations in FRB 20240114A with $\tau=1.5$ or 6.0 ms, obtained when plotted as shown in Fig.~\ref{fig:ML} becomes respectively $L=63.2-92.5$ MeV with $\tau=1.5$ ms and $L=52.6-106.5$ MeV with $\tau=6.0$~ms.

\begin{table*}
\caption{Same as Table~\ref{tab:NS_mass}, but with $\tau=1.5$ ms. The values in the bracket next to each entry indicate the relative deviation from the value when $\tau=3.0$~ms (c.f. Table~\ref{tab:NS_mass}).} 
\label{tab:NS_mass_tau15}
\begin {center}
\setlength{\tabcolsep}{5pt}
\begin{tabular}{ccccccc}
\hline\hline
QPO (Hz)  & $N_s/N_d$ & $R$ (km) & $M_{\rm max}/M_\odot$ & $M_{\rm min}/M_\odot$ & $L_{\rm max}$ (MeV) & $L_{\rm min}$ (MeV) \\ 
\hline
567.7   &  0 &  14  & 1.50 (3.02\%) & 1.32 (2.08\%) & 73.4 (5.37\%) & 68.2 (6.02\%) \\
    &    &  12  & 1.24 (3.58\%) & 1.06 (2.13\%) & 92.4 (4.62\%) & 87.1 (5.09\%) \\
    &  1 &  14  & 1.49 (2.90\%) & 1.30 (2.10\%) & 75.6 (4.67\%) & 71.2 (5.05\%) \\
    &    &  12  & 1.22 (3.66\%) & 1.02 (2.36\%) & 92.5 (4.08\%) & 88.3 (4.31\%) \\
\hline
655.5   &  0 & 14  & 1.71 (2.81\%) & 1.52 (2.09\%) & 68.1 (5.60\%) & 63.2 (6.20\%) \\
    &    & 12   & 1.41 (3.32\%) & 1.22 (2.22\%) & 87.3 (4.64\%) & 82.4 (5.11\%) \\
    &  1 & 14   & 1.71 (2.69\%) & 1.51 (1.99\%) & 70.9 (4.84\%) & 66.7 (5.24\%) \\
    &    & 12   & 1.39 (3.27\%) & 1.19 (2.30\%) & 87.9 (4.11\%) & 84.0 (4.36\%) \\
\hline \hline
\end{tabular}
\end {center}
\end{table*}

\begin{table*}
\caption{Same as Table~\ref{tab:NS_mass}, but with $\tau=6.0$ ms. The values in the bracket next to each entry indicate the relative deviation from the value when $\tau=3.0$~ms (c.f. Table~\ref{tab:NS_mass}).} 
\label{tab:NS_mass_tau60}
\begin {center}
\setlength{\tabcolsep}{5pt}
\begin{tabular}{ccccccc}
\hline\hline
QPO (Hz)  & $N_s/N_d$ & $R$ (km) & $M_{\rm max}/M_\odot$ & $M_{\rm min}/M_\odot$ & $L_{\rm max}$ (MeV) & $L_{\rm min}$ (MeV) \\ 
\hline
567.7   &  0 &  14  & 1.66 (7.13\%) & 1.25 (3.45\%) & 86.5 (11.6\%) & 57.0 (11.3\%) \\
    &    &  12  & 1.40 (8.86\%) & 1.00 (3.17\%) & 106.5 (10.0\%) & 74.9 (9.62\%) \\
    &  1 &  14  & 1.64 (6.78\%) & 1.23 (3.57\%) & 87.3 (10.1\%) & 61.3 (9.55\%) \\
    &    &  12  & 1.38 (9.01\%) & 0.961 (3.65\%) & 105.1 (8.95\%) & 77.8 (8.13\%) \\
\hline
655.5   &  0 & 14  & 2.08 (17.6\%) & 1.44 (3.65\%) & 77.7 (7.68\%) & 52.6 (11.6\%) \\
    &    & 12   & 1.57 (8.00\%) & 1.15 (3.58\%) & 100.6 (9.98\%) & 70.9 (9.65\%) \\
    &  1 & 14   & 1.86 (6.25\%) & 1.43 (3.52\%) & 82.3 (10.4\%) & 57.1 (9.89\%) \\
    &    & 12   & 1.55 (7.82\%) & 1.12 (3.80\%) & 99.8 (8.91\%) & 73.9 (8.20\%) \\
\hline \hline
\end{tabular}
\end {center}
\end{table*}

\bibliographystyle{aasjournalv7}
\bibliography{references}

@ARTICLE{2017LRR....20....7P,
       author = {{Paschalidis}, Vasileios and {Stergioulas}, Nikolaos},
        title = "{Rotating stars in relativity}",
      journal = {Living Reviews in Relativity},
         year = 2017,
        month = dec,
       volume = {20},
       number = {1},
          eid = {7},
        pages = {7},
          doi = {10.1007/s41114-017-0008-x},
archivePrefix = {arXiv},
       eprint = {1612.03050},
 primaryClass = {astro-ph.HE},
       adsurl = {https://ui.adsabs.harvard.edu/abs/2017LRR....20....7P}
}

@misc{wang2026gecamdiscoverysecondfrbassociated,
      title={GECAM discovery of the second FRB-associated Magnetar X-ray Burst from SGR J1935+2154}, 
      author={Chen-Wei Wang and Shao-Lin Xiong and Yue Wang and Wen-Jun Tan and Xiao-Bo Li and Dong-Zi Li and Yan-Qiu Zhang and Shu-Xu Yi and Ming-Yu Ge and Sheng-Lun Xie and Wang-Chen Xue and Bing Li and Cheng-Kui Li and Zheng-Hua An and Ce Cai and Pei-Yi Feng and Min Gao and Ke Gong and Dong-Ya Guo and Hao-Xuan Guo and Yue Huang and Jia-Cong Liu and Xin-Qiao Li and Ya-Qing Liu and Xiao-Jing Liu and Xiang Ma and Wen-Xi Peng and Rui Qiao and Yang-Zhao Ren and Li-Ming Song and Xi-Lei Sun and Jin Wang and Jin-Zhou Wang and Ping Wang and Xiang-Yang Wen and Shuo Xiao and Sheng Yang and Qi-Bin Yi and Zheng-Hang Yu and Da-Li Zhang and Fan Zhang and Wen-Long Zhang and Jin-Peng Zhang and Peng Zhang and Shuan-Nan Zhang and Zhen Zhang and Xiao-Yun Zhao and Yi Zhao and Chao Zheng and Shi-Jie Zheng},
      year={2026},
      eprint={2602.10895},
      archivePrefix={arXiv},
      primaryClass={astro-ph.HE},
      url={https://arxiv.org/abs/2602.10895}, 
}

@ARTICLE{2023Natur.613..253C,
       author = {{Chirenti}, Cecilia and {Dichiara}, Simone and {Lien}, Amy and {Miller}, M. Coleman and {Preece}, Robert},
        title = "{Kilohertz quasiperiodic oscillations in short gamma-ray bursts}",
      journal = {\nat},
         year = 2023,
        month = jan,
       volume = {613},
       number = {7943},
        pages = {253-256},
          doi = {10.1038/s41586-022-05497-0},
archivePrefix = {arXiv},
       eprint = {2301.02864},
 primaryClass = {astro-ph.HE},
       adsurl = {https://ui.adsabs.harvard.edu/abs/2023Natur.613..253C}
}

@ARTICLE{2025A&A...695L..12B,
       author = {{Bruni}, G. and {Piro}, L. and {Yang}, Y.-P. and {Palazzi}, E. and {Nicastro}, L. and {Rossi}, A. and {Savaglio}, S. and {Maiorano}, E. and {Zhang}, B.},
        title = "{Discovery of a persistent radio source associated with FRB 20240114A}",
      journal = {\aap},
         year = 2025,
        month = mar,
       volume = {695},
          eid = {L12},
        pages = {L12},
          doi = {10.1051/0004-6361/202453233},
       adsurl = {https://ui.adsabs.harvard.edu/abs/2025A&A...695L..12B}
}

@ARTICLE{2025ApJ...992L..35B,
       author = {{Bhardwaj}, M. and {Snelders}, M.~P. and {Hessels}, J.~W.~T. and {Gil de Paz}, A. and {Bhandari}, S. and {Marcote}, B. and {Kirichenko}, A. and {Ould-Boukattine}, O.~S. and {Kirsten}, F. and {Bempong-Manful}, E.~K. and {Bezrukovs}, V. and {Bray}, J.~D. and {Buttaccio}, S. and {Corongiu}, A. and {Feiler}, R. and {Gawro{\'n}ski}, M.~P. and {Giroletti}, M. and {Hewitt}, D.~M. and {Lindqvist}, M. and {Maccaferri}, G. and {Moroianu}, A. and {Nimmo}, K. and {Paragi}, Z. and {Puchalska}, W. and {Wang}, N. and {Williams-Baldwin}, D. and {Yuan}, J.~P.},
        title = "{A Hyperactive Fast Radio Burst Pinpointed in an SMC-like Satellite Host Galaxy}",
      journal = {\apjl},
         year = 2025,
        month = oct,
       volume = {992},
       number = {2},
          eid = {L35},
        pages = {L35},
          doi = {10.3847/2041-8213/ae0b68},
archivePrefix = {arXiv},
       eprint = {2506.11915},
 primaryClass = {astro-ph.HE},
       adsurl = {https://ui.adsabs.harvard.edu/abs/2025ApJ...992L..35B}
}

@ARTICLE{2024MNRAS.535.2449L,
       author = {{Lander}, S.~K.},
        title = "{The Meissner effect in neutron stars}",
      journal = {\mnras},
         year = 2024,
        month = dec,
       volume = {535},
       number = {3},
        pages = {2449-2468},
          doi = {10.1093/mnras/stae2453},
archivePrefix = {arXiv},
       eprint = {2411.08021},
 primaryClass = {astro-ph.HE},
       adsurl = {https://ui.adsabs.harvard.edu/abs/2024MNRAS.535.2449L}
}

@article{Lander_2026,
doi = {10.3847/2041-8213/ae31f5},
url = {https://doi.org/10.3847/2041-8213/ae31f5},
year = {2026},
month = {jan},
publisher = {The American Astronomical Society},
volume = {997},
number = {1},
pages = {L7},
author = {Lander, S. K. and Gourgouliatos, K. N. and Wadiasingh, Z. and Antonopoulou, D.},
title = {Signatures of Magnetic Flux Expulsion from Neutron Star Cores},
journal = {The Astrophysical Journal Letters}
}

@ARTICLE{2011IJMPD..20..989N,
       author = {{Nan}, Rendong and {Li}, Di and {Jin}, Chengjin and {Wang}, Qiming and {Zhu}, Lichun and {Zhu}, Wenbai and {Zhang}, Haiyan and {Yue}, Youling and {Qian}, Lei},
        title = "{The Five-Hundred Aperture Spherical Radio Telescope (fast) Project}",
      journal = {International Journal of Modern Physics D},
         year = 2011,
        month = jan,
       volume = {20},
       number = {6},
        pages = {989-1024},
          doi = {10.1142/S0218271811019335},
archivePrefix = {arXiv},
       eprint = {1105.3794},
 primaryClass = {astro-ph.IM},
       adsurl = {https://ui.adsabs.harvard.edu/abs/2011IJMPD..20..989N}
}

@misc{xiao2026evidencedampedmillisecondquasiperiodic,
      title={Evidence for a Damped Millisecond Quasi-Periodic Structure in a Fast Radio Burst}, 
      author={Shuo Xiao and Zheng-Huo Jiang and Di Li},
      year={2026},
      eprint={2601.03950},
      archivePrefix={arXiv},
      primaryClass={astro-ph.HE},
      url={https://arxiv.org/abs/2601.03950}, 
}

@ARTICLE{2021MNRAS.504.5880B,
       author = {{Bretz}, Joseph and {van Eysden}, C.~A. and {Link}, Bennett},
        title = "{Tangled magnetic field model of QPOs}",
      journal = {\mnras},
         year = 2021,
        month = jul,
       volume = {504},
       number = {4},
        pages = {5880-5898},
          doi = {10.1093/mnras/stab1220},
archivePrefix = {arXiv},
       eprint = {2206.09544},
 primaryClass = {astro-ph.HE},
       adsurl = {https://ui.adsabs.harvard.edu/abs/2021MNRAS.504.5880B}
}

@ARTICLE{2025arXiv250505373S,
       author = {{Suvorov}, Arthur G. and {Dehman}, Clara and {Pons}, Jos{\'e} A.},
        title = "{Late-blooming magnetars: awakening as ultra-long period objects after a dormant cooling epoch}",
      journal = {arXiv e-prints},
         year = 2025,
        month = may,
          eid = {arXiv:2505.05373},
        pages = {arXiv:2505.05373},
          doi = {10.48550/arXiv.2505.05373},
archivePrefix = {arXiv},
       eprint = {2505.05373},
 primaryClass = {astro-ph.HE},
       adsurl = {https://ui.adsabs.harvard.edu/abs/2025arXiv250505373S}
}

@ARTICLE{2016ApJS..224....6L,
       author = {{Link}, Bennett and {van Eysden}, C. Anthony},
        title = "{Supplement to {\textquotedblleft}Torsional Oscillations of a Magnetar with a Tangled Magnetic Field{\textquotedblright} (2016, ApJL, 823, L1)}",
      journal = {\apjs},
         year = 2016,
        month = may,
       volume = {224},
       number = {1},
          eid = {6},
        pages = {6},
          doi = {10.3847/0067-0049/224/1/6},
       adsurl = {https://ui.adsabs.harvard.edu/abs/2016ApJS..224....6L}
}

@INPROCEEDINGS{2017APS..APR.Y4007B,
       author = {{Bretz}, Joseph and {van Eysden}, Anthony and {Link}, Bennett},
        title = "{Preferential Excitation of Stellar Oscillations of a Magnetar with a Tangled Magnetic Field}",
    booktitle = {APS April Meeting Abstracts},
         year = 2017,
       series = {APS Meeting Abstracts},
       volume = {2017},
        month = jan,
          eid = {Y4.007},
        pages = {Y4.007},
       adsurl = {https://ui.adsabs.harvard.edu/abs/2017APS..APR.Y4007B}
}

@ARTICLE{2016ApJ...823L...1L,
       author = {{Link}, Bennett and {van Eysden}, C. Anthony},
        title = "{Torsional Oscillations of a Magnetar with a Tangled Magnetic Field}",
      journal = {\apjl},
         year = 2016,
        month = may,
       volume = {823},
       number = {1},
          eid = {L1},
        pages = {L1},
          doi = {10.3847/2041-8205/823/1/L1},
archivePrefix = {arXiv},
       eprint = {1604.02372},
 primaryClass = {astro-ph.HE},
       adsurl = {https://ui.adsabs.harvard.edu/abs/2016ApJ...823L...1L}
}

@INPROCEEDINGS{2019BAAS...51g.255H,
       author = {{Hallinan}, Gregg and {Ravi}, V. and {Weinreb}, S. and {Kocz}, J. and {Huang}, Y. and {Woody}, D.~P. and {Lamb}, J. and {D'Addario}, L. and {Catha}, M. and {Law}, C. and {Kulkarni}, S.~R. and {Phinney}, E.~S. and {Eastwood}, M.~W. and {Bouman}, K. and {McLaughlin}, M. and {Ransom}, S. and {Siemens}, X. and {Cordes}, J. and {Lynch}, R. and {Kaplan}, D. and {Brazier}, A. and {Bhatnagar}, S. and {Myers}, S. and {Walter}, F. and {Gaensler}, B.},
        title = "{The DSA-2000 {\textemdash} A Radio Survey Camera}",
    booktitle = {Bulletin of the American Astronomical Society},
         year = 2019,
       volume = {51},
        month = sep,
          eid = {255},
        pages = {255},
          doi = {10.48550/arXiv.1907.07648},
archivePrefix = {arXiv},
       eprint = {1907.07648},
 primaryClass = {astro-ph.IM},
       adsurl = {https://ui.adsabs.harvard.edu/abs/2019BAAS...51g.255H}
}

@ARTICLE{2022MNRAS.515.3698C,
       author = {{Crawford}, F. and {Hisano}, S. and {Golden}, M. and {Kikunaga}, T. and {Laity}, A. and {Zoeller}, D.},
        title = "{Four new fast radio bursts discovered in the Parkes 70-cm pulsar survey archive}",
      journal = {\mnras},
         year = 2022,
        month = sep,
       volume = {515},
       number = {3},
        pages = {3698-3702},
          doi = {10.1093/mnras/stac2101},
archivePrefix = {arXiv},
       eprint = {2207.12332},
 primaryClass = {astro-ph.HE},
       adsurl = {https://ui.adsabs.harvard.edu/abs/2022MNRAS.515.3698C}
}

@INPROCEEDINGS{2024AAS...24326104S,
       author = {{Sherman}, Myles and {Connor}, Liam and {Ravi}, Vikram and {Law}, Casey and {DSA-2000 Collaboration}},
        title = "{The DSA-2000 Fast Time Domain Search for Pulsars and Fast Radio Bursts}",
    booktitle = {American Astronomical Society Meeting Abstracts \#243},
         year = 2024,
       series = {American Astronomical Society Meeting Abstracts},
       volume = {243},
        month = feb,
          eid = {261.04},
        pages = {261.04},
       adsurl = {https://ui.adsabs.harvard.edu/abs/2024AAS...24326104S}
}

@INPROCEEDINGS{2019clrp.2020...28V,
       author = {{Vanderlinde}, Keith and {Liu}, Adrian and {Gaensler}, Bryan and {Bond}, Dick and {Hinshaw}, Gary and {Ng}, Cherry and {Chiang}, Cynthia and {Stairs}, Ingrid and {Brown}, Jo-Anne and {Sievers}, Jonathan and {Mena}, Juan and {Smith}, Kendrick and {Bandura}, Kevin and {Masui}, Kiyoshi and {Spekkens}, Kristine and {Belostotski}, Leo and {Dobbs}, Matt and {Turok}, Neil and {Boyle}, Patrick and {Rupen}, Michael and {Landecker}, Tom and {Pen}, Ue-Li and {Kaspi}, Victoria},
        title = "{The Canadian Hydrogen Observatory and Radio-transient Detector (CHORD)}",
    booktitle = {Canadian Long Range Plan for Astronomy and Astrophysics White Papers},
         year = 2019,
       volume = {2020},
        month = oct,
          eid = {28},
        pages = {28},
          doi = {10.5281/zenodo.3765414},
archivePrefix = {arXiv},
       eprint = {1911.01777},
 primaryClass = {astro-ph.IM},
       adsurl = {https://ui.adsabs.harvard.edu/abs/2019clrp.2020...28V}
}

@ARTICLE{2021ApJ...923....1P,
       author = {{Pleunis}, Ziggy and {Good}, Deborah C. and {Kaspi}, Victoria M. and {Mckinven}, Ryan and {Ransom}, Scott M. and {Scholz}, Paul and {Bandura}, Kevin and {Bhardwaj}, Mohit and {Boyle}, P.~J. and {Brar}, Charanjot and {Cassanelli}, Tomas and {Chawla}, Pragya and {(Adam) Dong}, Fengqiu and {Fonseca}, Emmanuel and {Gaensler}, B.~M. and {Josephy}, Alexander and {Kaczmarek}, Jane F. and {Leung}, Calvin and {Lin}, Hsiu-Hsien and {Masui}, Kiyoshi W. and {Mena-Parra}, Juan and {Michilli}, Daniele and {Ng}, Cherry and {Patel}, Chitrang and {Rafiei-Ravandi}, Masoud and {Rahman}, Mubdi and {Sanghavi}, Pranav and {Shin}, Kaitlyn and {Smith}, Kendrick M. and {Stairs}, Ingrid H. and {Tendulkar}, Shriharsh P.},
        title = "{Fast Radio Burst Morphology in the First CHIME/FRB Catalog}",
      journal = {\apj},
         year = 2021,
        month = dec,
       volume = {923},
       number = {1},
          eid = {1},
        pages = {1},
          doi = {10.3847/1538-4357/ac33ac},
archivePrefix = {arXiv},
       eprint = {2106.04356},
 primaryClass = {astro-ph.HE},
       adsurl = {https://ui.adsabs.harvard.edu/abs/2021ApJ...923....1P}
}

@ARTICLE{2025A&A...696A.194B,
       author = {{Bilous}, A.~V. and {van Leeuwen}, J. and {Maan}, Y. and {Pastor-Marazuela}, I. and {Oostrum}, L.~C. and {Rajwade}, K.~M. and {Wang}, Y.~Y.},
        title = "{An activity transition in FRB 20201124A: Methodological rigor, detection of frequency-dependent cessation, and a geometric magnetar model}",
      journal = {\aap},
         year = 2025,
        month = apr,
       volume = {696},
          eid = {A194},
        pages = {A194},
          doi = {10.1051/0004-6361/202451413},
archivePrefix = {arXiv},
       eprint = {2407.05366},
 primaryClass = {astro-ph.HE},
       adsurl = {https://ui.adsabs.harvard.edu/abs/2025A&A...696A.194B}
}

@ARTICLE{2023A&A...678A.149P,
       author = {{Pastor-Marazuela}, In{\'e}s and {van Leeuwen}, Joeri and {Bilous}, Anna and {Connor}, Liam and {Maan}, Yogesh and {Oostrum}, Leon and {Petroff}, Emily and {Straal}, Samayra and {Vohl}, Dany and {Adams}, Elizabeth A.~K. and {Adebahr}, Bj{\"o}rn and {Attema}, Jisk and {Boersma}, Oliver M. and {van den Brink}, R. and {van Cappellen}, W.~A. and {Coolen}, Arthur H.~W.~M. and {Damstra}, Sieds and {D{\'e}nes}, Helga and {Hess}, Kelley M. and {van der Hulst}, J.~M. and {Hut}, Boudewijn and {Kutkin}, Alexander and {Marcel Loose}, G. and {Lucero}, Danielle M. and {Mika}, {\'A}gnes and {Moss}, Vanessa A. and {Mulder}, Henk and {Norden}, Menno J. and {Oosterloo}, Tom A. and {Rajwade}, Kaustubh and {van der Schuur}, Daniel and {Sclocco}, Alessio and {Smits}, R. and {Ziemke}, Jacob},
        title = "{A fast radio burst with submillisecond quasi-periodic structure}",
      journal = {\aap},
         year = 2023,
        month = oct,
       volume = {678},
          eid = {A149},
        pages = {A149},
          doi = {10.1051/0004-6361/202243339},
archivePrefix = {arXiv},
       eprint = {2202.08002},
 primaryClass = {astro-ph.HE},
       adsurl = {https://ui.adsabs.harvard.edu/abs/2023A&A...678A.149P}
}

@ARTICLE{2022MNRAS.517.5080W,
       author = {{Wang}, Wei-Yang and {Jiang}, Jin-Chen and {Lee}, Kejia and {Xu}, Renxin and {Zhang}, Bing},
        title = "{Polarization of magnetospheric curvature radiation in repeating fast radio bursts}",
      journal = {\mnras},
         year = 2022,
        month = dec,
       volume = {517},
       number = {4},
        pages = {5080-5089},
          doi = {10.1093/mnras/stac3070},
archivePrefix = {arXiv},
       eprint = {2210.04401},
 primaryClass = {astro-ph.HE},
       adsurl = {https://ui.adsabs.harvard.edu/abs/2022MNRAS.517.5080W}
}

@ARTICLE{2022Natur.607..256C,
       author = {{CHIME/FRB Collaboration} and {Andersen}, Bridget C. and {Bandura}, Kevin and {Bhardwaj}, Mohit and {Boyle}, P.~J. and {Brar}, Charanjot and {Breitman}, Daniela and {Cassanelli}, Tomas and {Chatterjee}, Shami and {Chawla}, Pragya and {Cliche}, Jean-Fran{\c{c}}ois and {Cubranic}, Davor and {Curtin}, Alice P. and {Deng}, Meiling and {Dobbs}, Matt and {Dong}, Fengqiu Adam and {Fonseca}, Emmanuel and {Gaensler}, B.~M. and {Giri}, Utkarsh and {Good}, Deborah C. and {Hill}, Alex S. and {Josephy}, Alexander and {Kaczmarek}, J.~F. and {Kader}, Zarif and {Kania}, Joseph and {Kaspi}, Victoria M. and {Leung}, Calvin and {Li}, D.~Z. and {Lin}, Hsiu-Hsien and {Masui}, Kiyoshi W. and {McKinven}, Ryan and {Mena-Parra}, Juan and {Merryfield}, Marcus and {Meyers}, B.~W. and {Michilli}, D. and {Naidu}, Arun and {Newburgh}, Laura and {Ng}, C. and {Ordog}, Anna and {Patel}, Chitrang and {Pearlman}, Aaron B. and {Pen}, Ue-Li and {Petroff}, Emily and {Pleunis}, Ziggy and {Rafiei-Ravandi}, Masoud and {Rahman}, Mubdi and {Ransom}, Scott and {Renard}, Andre and {Sanghavi}, Pranav and {Scholz}, Paul and {Shaw}, J. Richard and {Shin}, Kaitlyn and {Siegel}, Seth R. and {Singh}, Saurabh and {Smith}, Kendrick and {Stairs}, Ingrid and {Tan}, Chia Min and {Tendulkar}, Shriharsh P. and {Vanderlinde}, Keith and {Wiebe}, D.~V. and {Wulf}, Dallas and {Zwaniga}, Andrew},
        title = "{Sub-second periodicity in a fast radio burst}",
      journal = {\nat},
         year = 2022,
        month = jul,
       volume = {607},
       number = {7918},
        pages = {256-259},
          doi = {10.1038/s41586-022-04841-8},
archivePrefix = {arXiv},
       eprint = {2107.08463},
 primaryClass = {astro-ph.HE},
       adsurl = {https://ui.adsabs.harvard.edu/abs/2022Natur.607..256C}
}

@ARTICLE{2006ApJ...653..593S,
       author = {{Strohmayer}, Tod E. and {Watts}, Anna L.},
        title = "{The 2004 Hyperflare from SGR 1806-20: Further Evidence for Global Torsional Vibrations}",
      journal = {\apj},
         year = 2006,
        month = dec,
       volume = {653},
       number = {1},
        pages = {593-601},
          doi = {10.1086/508703},
archivePrefix = {arXiv},
       eprint = {astro-ph/0608463},
 primaryClass = {astro-ph},
       adsurl = {https://ui.adsabs.harvard.edu/abs/2006ApJ...653..593S}
}

@ARTICLE{2006ApJ...637L.117W,
       author = {{Watts}, Anna L. and {Strohmayer}, Tod E.},
        title = "{Detection with RHESSI of High-Frequency X-Ray Oscillations in the Tailof the 2004 Hyperflare from SGR 1806-20}",
      journal = {\apjl},
         year = 2006,
        month = feb,
       volume = {637},
       number = {2},
        pages = {L117-L120},
          doi = {10.1086/500735},
archivePrefix = {arXiv},
       eprint = {astro-ph/0512630},
 primaryClass = {astro-ph},
       adsurl = {https://ui.adsabs.harvard.edu/abs/2006ApJ...637L.117W}
}

@ARTICLE{2005ApJ...632L.111S,
       author = {{Strohmayer}, Tod E. and {Watts}, Anna L.},
        title = "{Discovery of Fast X-Ray Oscillations during the 1998 Giant Flare from SGR 1900+14}",
      journal = {\apjl},
         year = 2005,
        month = oct,
       volume = {632},
       number = {2},
        pages = {L111-L114},
          doi = {10.1086/497911},
archivePrefix = {arXiv},
       eprint = {astro-ph/0508206},
 primaryClass = {astro-ph},
       adsurl = {https://ui.adsabs.harvard.edu/abs/2005ApJ...632L.111S}
}

@ARTICLE{2007AdSpR..40.1446W,
       author = {{Watts}, Anna L. and {Strohmayer}, Tod E.},
        title = "{Neutron star oscillations and QPOs during magnetar flares}",
      journal = {Advances in Space Research},
         year = 2007,
        month = jan,
       volume = {40},
       number = {10},
        pages = {1446-1452},
          doi = {10.1016/j.asr.2006.12.021},
archivePrefix = {arXiv},
       eprint = {astro-ph/0612252},
 primaryClass = {astro-ph},
       adsurl = {https://ui.adsabs.harvard.edu/abs/2007AdSpR..40.1446W}
}

@ARTICLE{2019ApJ...876L..23H,
       author = {{Hessels}, J.~W.~T. and {Spitler}, L.~G. and {Seymour}, A.~D. and {Cordes}, J.~M. and {Michilli}, D. and {Lynch}, R.~S. and {Gourdji}, K. and {Archibald}, A.~M. and {Bassa}, C.~G. and {Bower}, G.~C. and {Chatterjee}, S. and {Connor}, L. and {Crawford}, F. and {Deneva}, J.~S. and {Gajjar}, V. and {Kaspi}, V.~M. and {Keimpema}, A. and {Law}, C.~J. and {Marcote}, B. and {McLaughlin}, M.~A. and {Paragi}, Z. and {Petroff}, E. and {Ransom}, S.~M. and {Scholz}, P. and {Stappers}, B.~W. and {Tendulkar}, S.~P.},
        title = "{FRB 121102 Bursts Show Complex Time-Frequency Structure}",
      journal = {\apjl},
         year = 2019,
        month = may,
       volume = {876},
       number = {2},
          eid = {L23},
        pages = {L23},
          doi = {10.3847/2041-8213/ab13ae},
archivePrefix = {arXiv},
       eprint = {1811.10748},
 primaryClass = {astro-ph.HE},
       adsurl = {https://ui.adsabs.harvard.edu/abs/2019ApJ...876L..23H}
}

@ARTICLE{2025MNRAS.536.3220D,
       author = {{Dial}, T. and {Deller}, A.~T. and {Uttarkar}, P.~A. and {Lower}, M.~E. and {Shannon}, R.~M. and {Gourdji}, Kelly and {Marnoch}, Lachlan and {Bera}, A. and {Ryder}, Stuart D. and {Glowacki}, Marcin and {Prochaska}, J. Xavier},
        title = "{FRB 20230708A, a quasi-periodic FRB with unique temporal-polarimetric morphology}",
      journal = {\mnras},
         year = 2025,
        month = feb,
       volume = {536},
       number = {4},
        pages = {3220-3231},
          doi = {10.1093/mnras/stae2756},
archivePrefix = {arXiv},
       eprint = {2412.11347},
 primaryClass = {astro-ph.HE},
       adsurl = {https://ui.adsabs.harvard.edu/abs/2025MNRAS.536.3220D}
}

@ARTICLE{2024MNRAS.52710425S,
       author = {{Sheikh}, Sofia Z. and {Farah}, Wael and {Pollak}, Alexander W. and {Siemion}, Andrew P.~V. and {Chamma}, Mohammed A. and {Cruz}, Luigi F. and {Davis}, Roy H. and {DeBoer}, David R. and {Gajjar}, Vishal and {Karn}, Phil and {Kittling}, Jamar and {Lu}, Wenbin and {Masters}, Mark and {Premnath}, Pranav and {Schoultz}, Sarah and {Shumaker}, Carol and {Singh}, Gurmehar and {Snodgrass}, Michael},
        title = "{Characterization of the repeating FRB 20220912A with the Allen Telescope Array}",
      journal = {\mnras},
         year = 2024,
        month = feb,
       volume = {527},
       number = {4},
        pages = {10425-10439},
          doi = {10.1093/mnras/stad3630},
archivePrefix = {arXiv},
       eprint = {2312.07756},
 primaryClass = {astro-ph.HE},
       adsurl = {https://ui.adsabs.harvard.edu/abs/2024MNRAS.52710425S}
}

@ARTICLE{2023MNRAS.519..666J,
       author = {{Jahns}, J.~N. and {Spitler}, L.~G. and {Nimmo}, K. and {Hewitt}, D.~M. and {Snelders}, M.~P. and {Seymour}, A. and {Hessels}, J.~W.~T. and {Gourdji}, K. and {Michilli}, D. and {Hilmarsson}, G.~H.},
        title = "{The FRB 20121102A November rain in 2018 observed with the Arecibo Telescope}",
      journal = {\mnras},
         year = 2023,
        month = feb,
       volume = {519},
       number = {1},
        pages = {666-687},
          doi = {10.1093/mnras/stac3446},
archivePrefix = {arXiv},
       eprint = {2202.05705},
 primaryClass = {astro-ph.HE},
       adsurl = {https://ui.adsabs.harvard.edu/abs/2023MNRAS.519..666J}
}

@ARTICLE{2022RAA....22l4004N,
       author = {{Niu}, Jia-Rui and {Zhu}, Wei-Wei and {Zhang}, Bing and {Yuan}, Mao and {Zhou}, De-Jiang and {Zhang}, Yong-Kun and {Jiang}, Jin-Chen and {Han}, J.~L. and {Li}, Di and {Lee}, Ke-Jia and {Wang}, Pei and {Feng}, Yi and {Li}, Dong-Zi and {Luo}, Rui and {Wang}, Fa-Yin and {Dai}, Zi-Gao and {Miao}, Chen-Chen and {Niu}, Chen-Hui and {Xu}, Heng and {Zhang}, Chun-Feng and {Wang}, Wei-Yang and {Wang}, Bo-Jun and {Xu}, Jiang-Wei},
        title = "{FAST Observations of an Extremely Active Episode of FRB 20201124A. IV. Spin Period Search}",
      journal = {Research in Astronomy and Astrophysics},
         year = 2022,
        month = dec,
       volume = {22},
       number = {12},
          eid = {124004},
        pages = {124004},
          doi = {10.1088/1674-4527/ac995d},
archivePrefix = {arXiv},
       eprint = {2210.03610},
 primaryClass = {astro-ph.HE},
       adsurl = {https://ui.adsabs.harvard.edu/abs/2022RAA....22l4004N}
}

@ARTICLE{2025ApJ...982...45B,
       author = {{Beniamini}, Paz and {Kumar}, Pawan},
        title = "{The Role of Magnetic and Rotation Axis Alignment in Driving Fast Radio Burst Phenomenology}",
      journal = {\apj},
         year = 2025,
        month = mar,
       volume = {982},
       number = {1},
          eid = {45},
        pages = {45},
          doi = {10.3847/1538-4357/adb8e6},
archivePrefix = {arXiv},
       eprint = {2410.19043},
 primaryClass = {astro-ph.HE},
       adsurl = {https://ui.adsabs.harvard.edu/abs/2025ApJ...982...45B}
}

@ARTICLE{2025arXiv251221889K,
       author = {{Kumar}, Ajay and {Maan}, Yogesh and {Lal}, Banshi and {Bhusare}, Yash and {Tendulkar}, Shriharsh P. and {Marthi}, Visweshwar Ram and {Majee}, Puja},
        title = "{Long-term monitoring of repeating FRB 20220912A with the uGMRT at low radio frequencies}",
      journal = {arXiv e-prints},
         year = 2025,
        month = dec,
          eid = {arXiv:2512.21889},
        pages = {arXiv:2512.21889},
          doi = {10.48550/arXiv.2512.21889},
archivePrefix = {arXiv},
       eprint = {2512.21889},
 primaryClass = {astro-ph.HE},
       adsurl = {https://ui.adsabs.harvard.edu/abs/2025arXiv251221889K}
}

@ARTICLE{2020ApJ...904L..21Y,
       author = {{Younes}, George and {G{\"u}ver}, Tolga and {Kouveliotou}, Chryssa and {Baring}, Matthew G. and {Hu}, Chin-Ping and {Wadiasingh}, Zorawar and {Begi{\c{c}}arslan}, Beste and {Enoto}, Teruaki and {G{\"o}{\u{g}}{\"u}{\c{s}}}, Ersin and {Lin}, Lin and {Harding}, Alice K. and {van der Horst}, Alexander J. and {Majid}, Walid A. and {Guillot}, Sebastien and {Malacaria}, Christian},
        title = "{NICER View of the 2020 Burst Storm and Persistent Emission of SGR 1935+2154}",
      journal = {\apjl},
         year = 2020,
        month = dec,
       volume = {904},
       number = {2},
          eid = {L21},
        pages = {L21},
          doi = {10.3847/2041-8213/abc94c},
archivePrefix = {arXiv},
       eprint = {2009.07886},
 primaryClass = {astro-ph.HE},
       adsurl = {https://ui.adsabs.harvard.edu/abs/2020ApJ...904L..21Y}
}

@ARTICLE{2024ApJ...972L..20N,
       author = {{Niu}, J.~R. and {Wang}, W.~Y. and {Jiang}, J.~C. and {Qu}, Y. and {Zhou}, D.~J. and {Zhu}, W.~W. and {Lee}, K.~J. and {Han}, J.~L. and {Zhang}, B. and {Li}, D. and {Cao}, S. and {Fang}, Z.~Y. and {Feng}, Y. and {Fu}, Q.~Y. and {Jiang}, P. and {Jing}, W.~C. and {Li}, J. and {Li}, Y. and {Luo}, R. and {Meng}, L.~Q. and {Miao}, C.~C. and {Miao}, X.~L. and {Niu}, C.~H. and {Pan}, Y.~C. and {Wang}, B.~J. and {Wang}, F.~Y. and {Wang}, H.~Z. and {Wang}, P. and {Wu}, Q. and {Wu}, Z.~W. and {Xu}, H. and {Xu}, J.~W. and {Xu}, L. and {Xue}, M.~Y. and {Yang}, Y.~P. and {Yuan}, M. and {Yue}, Y.~L. and {Zhao}, D. and {Zhang}, C.~F. and {Zhang}, D.~D. and {Zhang}, J.~S. and {Zhang}, S.~B. and {Zhang}, Y.~K. and {Zhu}, Y.~H.},
        title = "{Sudden Polarization Angle Jumps of the Repeating Fast Radio Burst FRB 20201124A}",
      journal = {\apjl},
         year = 2024,
        month = sep,
       volume = {972},
       number = {2},
          eid = {L20},
        pages = {L20},
          doi = {10.3847/2041-8213/ad7023},
archivePrefix = {arXiv},
       eprint = {2407.10540},
 primaryClass = {astro-ph.HE},
       adsurl = {https://ui.adsabs.harvard.edu/abs/2024ApJ...972L..20N}
}

@ARTICLE{2024MNRAS.533.2133C,
       author = {{Cooper}, A.~J. and {Wadiasingh}, Z.},
        title = "{Beyond the Rotational Deathline: Radio Emission from Ultra-long Period Magnetars}",
      journal = {\mnras},
         year = 2024,
        month = sep,
       volume = {533},
       number = {2},
        pages = {2133-2155},
          doi = {10.1093/mnras/stae1813},
archivePrefix = {arXiv},
       eprint = {2406.04135},
 primaryClass = {astro-ph.HE},
       adsurl = {https://ui.adsabs.harvard.edu/abs/2024MNRAS.533.2133C}
}

@ARTICLE{2023MNRAS.520.1872B,
       author = {{Beniamini}, P. and {Wadiasingh}, Z. and {Hare}, J. and {Rajwade}, K.~M. and {Younes}, G. and {van der Horst}, A.~J.},
        title = "{Evidence for an abundant old population of Galactic ultra-long period magnetars and implications for fast radio bursts}",
      journal = {\mnras},
         year = 2023,
        month = apr,
       volume = {520},
       number = {2},
        pages = {1872-1894},
          doi = {10.1093/mnras/stad208},
archivePrefix = {arXiv},
       eprint = {2210.09323},
 primaryClass = {astro-ph.HE},
       adsurl = {https://ui.adsabs.harvard.edu/abs/2023MNRAS.520.1872B}
}

@ARTICLE{2020ApJ...896..142B,
       author = {{Beloborodov}, Andrei M.},
        title = "{Blast Waves from Magnetar Flares and Fast Radio Bursts}",
      journal = {\apj},
         year = 2020,
        month = jun,
       volume = {896},
       number = {2},
          eid = {142},
        pages = {142},
          doi = {10.3847/1538-4357/ab83eb},
archivePrefix = {arXiv},
       eprint = {1908.07743},
 primaryClass = {astro-ph.HE},
       adsurl = {https://ui.adsabs.harvard.edu/abs/2020ApJ...896..142B}
}

@ARTICLE{2017MNRAS.468.2726K,
       author = {{Kumar}, Pawan and {Lu}, Wenbin and {Bhattacharya}, Mukul},
        title = "{Fast radio burst source properties and curvature radiation model}",
      journal = {\mnras},
         year = 2017,
        month = jul,
       volume = {468},
       number = {3},
        pages = {2726-2739},
          doi = {10.1093/mnras/stx665},
archivePrefix = {arXiv},
       eprint = {1703.06139},
 primaryClass = {astro-ph.HE},
       adsurl = {https://ui.adsabs.harvard.edu/abs/2017MNRAS.468.2726K}
}

@ARTICLE{2025Natur.637...48N,
       author = {{Nimmo}, Kenzie and {Pleunis}, Ziggy and {Beniamini}, Paz and {Kumar}, Pawan and {Lanman}, Adam E. and {Li}, D.~Z. and {Main}, Robert and {Sammons}, Mawson W. and {Andrew}, Shion and {Bhardwaj}, Mohit and {Chatterjee}, Shami and {Curtin}, Alice P. and {Fonseca}, Emmanuel and {Gaensler}, B.~M. and {Joseph}, Ronniy C. and {Kader}, Zarif and {Kaspi}, Victoria M. and {Lazda}, Mattias and {Leung}, Calvin and {Masui}, Kiyoshi W. and {Mckinven}, Ryan and {Michilli}, Daniele and {Pandhi}, Ayush and {Pearlman}, Aaron B. and {Rafiei-Ravandi}, Masoud and {Sand}, Ketan R. and {Shin}, Kaitlyn and {Smith}, Kendrick and {Stairs}, Ingrid H.},
        title = "{Magnetospheric origin of a fast radio burst constrained using scintillation}",
      journal = {\nat},
         year = 2025,
        month = jan,
       volume = {637},
       number = {8044},
        pages = {48-51},
          doi = {10.1038/s41586-024-08297-w},
archivePrefix = {arXiv},
       eprint = {2406.11053},
 primaryClass = {astro-ph.HE},
       adsurl = {https://ui.adsabs.harvard.edu/abs/2025Natur.637...48N}
}

@ARTICLE{2021ApJ...922..166L,
       author = {{Lyutikov}, Maxim},
        title = "{Coherent Emission in Pulsars, Magnetars, and Fast Radio Bursts: Reconnection-driven Free Electron Laser}",
      journal = {\apj},
         year = 2021,
        month = dec,
       volume = {922},
       number = {2},
          eid = {166},
        pages = {166},
          doi = {10.3847/1538-4357/ac1b32},
archivePrefix = {arXiv},
       eprint = {2102.07010},
 primaryClass = {astro-ph.HE},
       adsurl = {https://ui.adsabs.harvard.edu/abs/2021ApJ...922..166L}
}

@ARTICLE{2025ApJ...983L..16Y,
       author = {{Yamasaki}, Shotaro and {Totani}, Tomonori},
        title = "{Time{\textendash}Frequency Correlation of Repeating Fast Radio Bursts: Correlated Aftershocks Tend to Exhibit Downward Frequency Drifts}",
      journal = {\apjl},
         year = 2025,
        month = apr,
       volume = {983},
       number = {1},
          eid = {L16},
        pages = {L16},
          doi = {10.3847/2041-8213/adc10b},
archivePrefix = {arXiv},
       eprint = {2412.04313},
 primaryClass = {astro-ph.HE},
       adsurl = {https://ui.adsabs.harvard.edu/abs/2025ApJ...983L..16Y}
}

@ARTICLE{2025Natur.637...43M,
       author = {{Mckinven}, Ryan and {Bhardwaj}, Mohit and {Eftekhari}, Tarraneh and {Kilpatrick}, Charles D. and {Kirichenko}, Aida and {Pal}, Arpan and {Cook}, Amanda M. and {Gaensler}, B.~M. and {Giri}, Utkarsh and {Kaspi}, Victoria M. and {Michilli}, Daniele and {Nimmo}, Kenzie and {Pearlman}, Aaron B. and {Pleunis}, Ziggy and {Sand}, Ketan R. and {Stairs}, Ingrid and {Andersen}, Bridget C. and {Andrew}, Shion and {Bandura}, Kevin and {Brar}, Charanjot and {Cassanelli}, Tomas and {Chatterjee}, Shami and {Curtin}, Alice P. and {Dong}, Fengqiu Adam and {Eadie}, Gwendolyn and {Fonseca}, Emmanuel and {Ibik}, Adaeze L. and {Kaczmarek}, Jane F. and {Kharel}, Bikash and {Lazda}, Mattias and {Leung}, Calvin and {Li}, Dongzi and {Main}, Robert and {Masui}, Kiyoshi W. and {Mena-Parra}, Juan and {Ng}, Cherry and {Pandhi}, Ayush and {Patil}, Swarali Shivraj and {Prochaska}, J. Xavier and {Rafiei-Ravandi}, Masoud and {Scholz}, Paul and {Shah}, Vishwangi and {Shin}, Kaitlyn and {Smith}, Kendrick},
        title = "{A pulsar-like polarization angle swing from a nearby fast radio burst}",
      journal = {\nat},
         year = 2025,
        month = jan,
       volume = {637},
       number = {8044},
        pages = {43-47},
          doi = {10.1038/s41586-024-08184-4},
       adsurl = {https://ui.adsabs.harvard.edu/abs/2025Natur.637...43M}
}

@ARTICLE{2025A&A...702A.248B,
       author = {{Bethapudi}, S. and {Li}, D.~Z. and {Spitler}, L.~G. and {Marthi}, V.~R. and {Bause}, M.~L. and {Main}, R.~A. and {Wharton}, R.~S.},
        title = "{Constraining the origin of the long-term periodicity of FRB 20180916B with polarization position angle}",
      journal = {\aap},
         year = 2025,
        month = oct,
       volume = {702},
          eid = {A248},
        pages = {A248},
          doi = {10.1051/0004-6361/202556347},
archivePrefix = {arXiv},
       eprint = {2507.07651},
 primaryClass = {astro-ph.HE},
       adsurl = {https://ui.adsabs.harvard.edu/abs/2025A&A...702A.248B}
}

@ARTICLE{2025arXiv251224936K,
       author = {{Katz}, J.~I.},
        title = "{Searching for Periodicity in FRB 20240114A}",
      journal = {arXiv e-prints},
         year = 2025,
        month = dec,
          eid = {arXiv:2512.24936},
        pages = {arXiv:2512.24936},
          doi = {10.48550/arXiv.2512.24936},
archivePrefix = {arXiv},
       eprint = {2512.24936},
 primaryClass = {astro-ph.HE},
       adsurl = {https://ui.adsabs.harvard.edu/abs/2025arXiv251224936K}
}

@ARTICLE{2015ApJ...810...66H,
       author = {{Huppenkothen}, Daniela and {Brewer}, Brendon J. and {Hogg}, David W. and {Murray}, Iain and {Frean}, Marcus and {Elenbaas}, Chris and {Watts}, Anna L. and {Levin}, Yuri and {van der Horst}, Alexander J. and {Kouveliotou}, Chryssa},
        title = "{Dissecting Magnetar Variability with Bayesian Hierarchical Models}",
      journal = {\apj},
         year = 2015,
        month = sep,
       volume = {810},
       number = {1},
          eid = {66},
        pages = {66},
          doi = {10.1088/0004-637X/810/1/66},
archivePrefix = {arXiv},
       eprint = {1501.05251},
 primaryClass = {astro-ph.HE},
       adsurl = {https://ui.adsabs.harvard.edu/abs/2015ApJ...810...66H}
}

@ARTICLE{2014ApJ...795..114H,
       author = {{Huppenkothen}, D. and {Heil}, L.~M. and {Watts}, A.~L. and {G{\"o}{\u{g}}{\"u}{\c{s}}}, E.},
        title = "{Quasi-periodic Oscillations in Short Recurring Bursts of Magnetars SGR 1806-20 and SGR 1900+14 Observed with RXTE}",
      journal = {\apj},
         year = 2014,
        month = nov,
       volume = {795},
       number = {2},
          eid = {114},
        pages = {114},
          doi = {10.1088/0004-637X/795/2/114},
archivePrefix = {arXiv},
       eprint = {1409.7642},
 primaryClass = {astro-ph.HE},
       adsurl = {https://ui.adsabs.harvard.edu/abs/2014ApJ...795..114H}
}

@ARTICLE{2014ApJ...787..128H,
       author = {{Huppenkothen}, D. and {D'Angelo}, C. and {Watts}, A.~L. and {Heil}, L. and {van der Klis}, M. and {van der Horst}, A.~J. and {Kouveliotou}, C. and {Baring}, M.~G. and {G{\"o}{\u{g}}{\"u}{\c{s}}}, E. and {Granot}, J. and {Kaneko}, Y. and {Lin}, L. and {von Kienlin}, A. and {Younes}, G.},
        title = "{Quasi-periodic Oscillations in Short Recurring Bursts of the Soft Gamma Repeater J1550-5418}",
      journal = {\apj},
         year = 2014,
        month = jun,
       volume = {787},
       number = {2},
          eid = {128},
        pages = {128},
          doi = {10.1088/0004-637X/787/2/128},
archivePrefix = {arXiv},
       eprint = {1404.2756},
 primaryClass = {astro-ph.HE},
       adsurl = {https://ui.adsabs.harvard.edu/abs/2014ApJ...787..128H}
}

@ARTICLE{2025ApJ...980..211B,
       author = {{Beniamini}, Paz and {Wadiasingh}, Zorawar and {Trigg}, Aaron and {Chirenti}, Cecilia and {Burns}, Eric and {Younes}, George and {Negro}, Michela and {Granot}, Jonathan},
        title = "{Extragalactic Magnetar Giant Flares: Population Implications, Rates, and Prospects for Gamma-Rays, Gravitational Waves, and Neutrinos}",
      journal = {\apj},
         year = 2025,
        month = feb,
       volume = {980},
       number = {2},
          eid = {211},
        pages = {211},
          doi = {10.3847/1538-4357/ada947},
archivePrefix = {arXiv},
       eprint = {2411.16846},
 primaryClass = {astro-ph.HE},
       adsurl = {https://ui.adsabs.harvard.edu/abs/2025ApJ...980..211B}
}

@ARTICLE{2023MNRAS.526.2795T,
       author = {{Totani}, Tomonori and {Tsuzuki}, Yuya},
        title = "{Fast radio bursts trigger aftershocks resembling earthquakes, but not solar flares}",
      journal = {\mnras},
         year = 2023,
        month = dec,
       volume = {526},
       number = {2},
        pages = {2795-2811},
          doi = {10.1093/mnras/stad2532},
archivePrefix = {arXiv},
       eprint = {2306.13612},
 primaryClass = {astro-ph.HE},
       adsurl = {https://ui.adsabs.harvard.edu/abs/2023MNRAS.526.2795T}
}

@ARTICLE{2014MNRAS.442L...9L,
       author = {{Lyubarsky}, Yu.},
        title = "{A model for fast extragalactic radio bursts.}",
      journal = {\mnras},
         year = 2014,
        month = jul,
       volume = {442},
        pages = {L9-L13},
          doi = {10.1093/mnrasl/slu046},
archivePrefix = {arXiv},
       eprint = {1401.6674},
 primaryClass = {astro-ph.HE},
       adsurl = {https://ui.adsabs.harvard.edu/abs/2014MNRAS.442L...9L}
}

@ARTICLE{2017ApJ...843L..26B,
       author = {{Beloborodov}, Andrei M.},
        title = "{A Flaring Magnetar in FRB 121102?}",
      journal = {\apjl},
         year = 2017,
        month = jul,
       volume = {843},
       number = {2},
          eid = {L26},
        pages = {L26},
          doi = {10.3847/2041-8213/aa78f3},
archivePrefix = {arXiv},
       eprint = {1702.08644},
 primaryClass = {astro-ph.HE},
       adsurl = {https://ui.adsabs.harvard.edu/abs/2017ApJ...843L..26B}
}

@ARTICLE{2025ApJ...995L..57B,
       author = {{Burnaz}, Louis and {Most}, Elias R. and {Bransgrove}, Ashley},
        title = "{Crustal Quakes Spark Magnetospheric Blasts: Imprints of Realistic Magnetar Crust Oscillations on the Fast Radio Burst Signal}",
      journal = {\apjl},
         year = 2025,
        month = dec,
       volume = {995},
       number = {2},
          eid = {L57},
        pages = {L57},
          doi = {10.3847/2041-8213/ae2466},
archivePrefix = {arXiv},
       eprint = {2508.18033},
 primaryClass = {astro-ph.HE},
       adsurl = {https://ui.adsabs.harvard.edu/abs/2025ApJ...995L..57B}
}

@ARTICLE{2024Natur.626..500H,
       author = {{Hu}, Chin-Ping and {Narita}, Takuto and {Enoto}, Teruaki and {Younes}, George and {Wadiasingh}, Zorawar and {Baring}, Matthew G. and {Ho}, Wynn C.~G. and {Guillot}, Sebastien and {Ray}, Paul S. and {G{\"u}ver}, Tolga and {Rajwade}, Kaustubh and {Arzoumanian}, Zaven and {Kouveliotou}, Chryssa and {Harding}, Alice K. and {Gendreau}, Keith C.},
        title = "{Rapid spin changes around a magnetar fast radio burst}",
      journal = {\nat},
         year = 2024,
        month = feb,
       volume = {626},
       number = {7999},
        pages = {500-504},
          doi = {10.1038/s41586-023-07012-5},
archivePrefix = {arXiv},
       eprint = {2402.09291},
 primaryClass = {astro-ph.HE},
       adsurl = {https://ui.adsabs.harvard.edu/abs/2024Natur.626..500H}
}

@ARTICLE{2025ApJ...989...63H,
       author = {{Hu}, Chin-Ping and {Wadiasingh}, Zorawar and {Ho}, Wynn C.~G. and {Baring}, Matthew G. and {Younes}, George A. and {Enoto}, Teruaki and {Guillot}, Sebastien and {G{\"u}ver}, Tolga and {Bause}, Marlon L. and {Stewart}, Rachael and {Van Kooten}, Alex and {Kouveliotou}, Chryssa},
        title = "{Rapid Spectral Evolution of SGR 1935+2154 during Its 2022 Outburst}",
      journal = {\apj},
         year = 2025,
        month = aug,
       volume = {989},
       number = {1},
          eid = {63},
        pages = {63},
          doi = {10.3847/1538-4357/adea4e},
archivePrefix = {arXiv},
       eprint = {2504.21615},
 primaryClass = {astro-ph.HE},
       adsurl = {https://ui.adsabs.harvard.edu/abs/2025ApJ...989...63H}
}

@ARTICLE{2019MNRAS.485.4091M,
       author = {{Metzger}, Brian D. and {Margalit}, Ben and {Sironi}, Lorenzo},
        title = "{Fast radio bursts as synchrotron maser emission from decelerating relativistic blast waves}",
      journal = {\mnras},
         year = 2019,
        month = may,
       volume = {485},
       number = {3},
        pages = {4091-4106},
          doi = {10.1093/mnras/stz700},
archivePrefix = {arXiv},
       eprint = {1902.01866},
 primaryClass = {astro-ph.HE},
       adsurl = {https://ui.adsabs.harvard.edu/abs/2019MNRAS.485.4091M}
}

@ARTICLE{2018ApJ...852..140W,
       author = {{Wang}, Weiyang and {Luo}, Rui and {Yue}, Han and {Chen}, Xuelei and {Lee}, Kejia and {Xu}, Renxin},
        title = "{FRB 121102: A Starquake-induced Repeater?}",
      journal = {\apj},
         year = 2018,
        month = jan,
       volume = {852},
       number = {2},
          eid = {140},
        pages = {140},
          doi = {10.3847/1538-4357/aaa025},
archivePrefix = {arXiv},
       eprint = {1710.00541},
 primaryClass = {astro-ph.HE},
       adsurl = {https://ui.adsabs.harvard.edu/abs/2018ApJ...852..140W}
}

@ARTICLE{2019MNRAS.488.5887S,
       author = {{Suvorov}, A.~G. and {Kokkotas}, K.~D.},
        title = "{Young magnetars with fracturing crusts as fast radio burst repeaters}",
      journal = {\mnras},
         year = 2019,
        month = oct,
       volume = {488},
       number = {4},
        pages = {5887-5897},
          doi = {10.1093/mnras/stz2052},
archivePrefix = {arXiv},
       eprint = {1907.10394},
 primaryClass = {astro-ph.HE},
       adsurl = {https://ui.adsabs.harvard.edu/abs/2019MNRAS.488.5887S}
}

@ARTICLE{2019A&ARv..27....4P,
       author = {{Petroff}, E. and {Hessels}, J.~W.~T. and {Lorimer}, D.~R.},
        title = "{Fast radio bursts}",
      journal = {\aapr},
         year = 2019,
        month = dec,
       volume = {27},
       number = {1},
          eid = {4},
        pages = {4},
          doi = {10.1007/s00159-019-0116-6},
archivePrefix = {arXiv},
       eprint = {1904.07947},
 primaryClass = {astro-ph.HE},
       adsurl = {https://ui.adsabs.harvard.edu/abs/2019A&ARv..27....4P}
}

@ARTICLE{2022ApJ...931...56L,
       author = {{Li}, Xiaobo and {Ge}, Mingyu and {Lin}, Lin and {Zhang}, Shuang-Nan and {Song}, Liming and {Cao}, Xuelei and {Zhang}, Bing and {Lu}, Fangjun and {Xu}, Yupeng and {Xiong}, Shaolin and {Tuo}, Youli and {Tan}, Ying and {Jiang}, Weichun and {Qu}, Jinlu and {Zhang}, Shu and {Wang}, Lingjun and {Wang}, Jieshuang and {Zhang}, Binbin and {Zhang}, Peng and {Li}, Chengkui and {Liu}, Congzhan and {Li}, Tipei and {Bu}, Qingcui and {Cai}, Ce and {Chen}, Yong and {Chen}, Yupeng and {Chang}, Zhi and {Chen}, Li and {Chen}, Tianxiang and {Chen}, Yibao and {Cui}, Weiwei and {Du}, Yuanyuan and {Gao}, Guanhua and {Gao}, He and {Gu}, Yudong and {Guan}, Ju and {Guo}, Chengcheng and {Han}, Dawei and {Huang}, Yue and {Huo}, Jia and {Jia}, Shumei and {Jin}, Jing and {Kong}, Lingda and {Li}, Bing and {Li}, Gang and {Li}, Wei and {Li}, Xian and {Li}, Xufang and {Li}, Zhengwei and {Liang}, Xiaohua and {Liao}, Jinyuan and {Liu}, Hexin and {Liu}, Hongwei and {Liu}, Xiaojing and {Lu}, Xuefeng and {Luo}, Qi and {Luo}, Tao and {Ma}, Binyuan and {Ma}, Ruican and {Ma}, Xiang and {Meng}, Bin and {Nang}, Yi and {Nie}, Jianyin and {Ou}, Ge and {Ren}, Xiaoqin and {Sai}, Na and {Song}, Xinying and {Sun}, Liang and {Tao}, Lian and {Wang}, Chen and {Wang}, Pengju and {Wang}, Wenshuai and {Wang}, Yusa and {Wen}, Xiangyang and {Wu}, Bobing and {Wu}, Baiyang and {Wu}, Mei and {Xiao}, Shuo and {Yang}, Sheng and {Yang}, Yanji and {Yi}, Qibin and {Yin}, Qianqing and {You}, Yuan and {Yu}, Wei and {Zhang}, Fan and {Zhang}, Hongmei and {Zhang}, Juan and {Zhang}, Wanchang and {Zhang}, Wei and {Zhang}, Yifei and {Zhang}, Yuanhang and {Zhao}, Haisheng and {Zhao}, Xiaofan and {Zheng}, Shijie and {Zhou}, Dengke},
        title = "{Quasi-periodic Oscillations of the X-Ray Burst from the Magnetar SGR J1935-2154 and Associated with the Fast Radio Burst FRB 200428}",
      journal = {\apj},
         year = 2022,
        month = may,
       volume = {931},
       number = {1},
          eid = {56},
        pages = {56},
          doi = {10.3847/1538-4357/ac6587},
archivePrefix = {arXiv},
       eprint = {2204.03253},
 primaryClass = {astro-ph.HE},
       adsurl = {https://ui.adsabs.harvard.edu/abs/2022ApJ...931...56L}
}

@INPROCEEDINGS{2021sf2a.conf..405V,
       author = {{Voisin}, G.},
        title = "{A maze in(g) FRB models}",
    booktitle = {SF2A-2021: Proceedings of the Annual meeting of the French Society of Astronomy and Astrophysics},
         year = 2021,
       editor = {{Siebert}, A. and {Bailli{\'e}}, K. and {Lagadec}, E. and {Lagarde}, N. and {Malzac}, J. and {Marquette}, J. -B. and {N'Diaye}, M. and {Richard}, J. and {Venot}, O.},
        month = dec,
        pages = {405-412},
       adsurl = {https://ui.adsabs.harvard.edu/abs/2021sf2a.conf..405V}
}

@ARTICLE{2021SCPMA..6449501X,
       author = {{Xiao}, Di and {Wang}, FaYin and {Dai}, ZiGao},
        title = "{The physics of fast radio bursts}",
      journal = {Science China Physics, Mechanics, and Astronomy},
         year = 2021,
        month = apr,
       volume = {64},
       number = {4},
          eid = {249501},
        pages = {249501},
          doi = {10.1007/s11433-020-1661-7},
archivePrefix = {arXiv},
       eprint = {2101.04907},
 primaryClass = {astro-ph.HE},
       adsurl = {https://ui.adsabs.harvard.edu/abs/2021SCPMA..6449501X}
}

@ARTICLE{2020MNRAS.498..651B,
       author = {{Beniamini}, Paz and {Kumar}, Pawan},
        title = "{What does FRB light-curve variability tell us about the emission mechanism?}",
      journal = {\mnras},
         year = 2020,
        month = oct,
       volume = {498},
       number = {1},
        pages = {651-664},
          doi = {10.1093/mnras/staa2489},
archivePrefix = {arXiv},
       eprint = {2007.07265},
 primaryClass = {astro-ph.HE},
       adsurl = {https://ui.adsabs.harvard.edu/abs/2020MNRAS.498..651B}
}

@ARTICLE{2021NatAs...5..372R,
       author = {{Ridnaia}, A. and {Svinkin}, D. and {Frederiks}, D. and {Bykov}, A. and {Popov}, S. and {Aptekar}, R. and {Golenetskii}, S. and {Lysenko}, A. and {Tsvetkova}, A. and {Ulanov}, M. and {Cline}, T.~L.},
        title = "{A peculiar hard X-ray counterpart of a Galactic fast radio burst}",
      journal = {Nature Astronomy},
         year = 2021,
        month = apr,
       volume = {5},
        pages = {372-377},
          doi = {10.1038/s41550-020-01265-0},
archivePrefix = {arXiv},
       eprint = {2005.11178},
 primaryClass = {astro-ph.HE},
       adsurl = {https://ui.adsabs.harvard.edu/abs/2021NatAs...5..372R}
}

@ARTICLE{2023ApJ...953...67G,
       author = {{Ge}, M.~Y. and {Liu}, C.~Z. and {Zhang}, S.~N. and {Lu}, F.~J. and {Zhang}, Z. and {Chang}, Z. and {Tuo}, Y.~L. and {Li}, X.~B. and {Li}, C.~K. and {Xiong}, S.~L. and {Cai}, C. and {Li}, X.~F. and {Zhang}, R. and {Dai}, Z.~G. and {Qu}, J.~L. and {Song}, L.~M. and {Zhang}, S. and {Wang}, L.~J.},
        title = "{Reanalysis of the X-Ray-burst-associated FRB 200428 with Insight-HXMT Observations}",
      journal = {\apj},
         year = 2023,
        month = aug,
       volume = {953},
       number = {1},
          eid = {67},
        pages = {67},
          doi = {10.3847/1538-4357/acda1d},
archivePrefix = {arXiv},
       eprint = {2302.00176},
 primaryClass = {astro-ph.HE},
       adsurl = {https://ui.adsabs.harvard.edu/abs/2023ApJ...953...67G}
}

@ARTICLE{2021NatAs...5..378L,
       author = {{Li}, C.~K. and {Lin}, L. and {Xiong}, S.~L. and {Ge}, M.~Y. and {Li}, X.~B. and {Li}, T.~P. and {Lu}, F.~J. and {Zhang}, S.~N. and {Tuo}, Y.~L. and {Nang}, Y. and {Zhang}, B. and {Xiao}, S. and {Chen}, Y. and {Song}, L.~M. and {Xu}, Y.~P. and {Liu}, C.~Z. and {Jia}, S.~M. and {Cao}, X.~L. and {Qu}, J.~L. and {Zhang}, S. and {Gu}, Y.~D. and {Liao}, J.~Y. and {Zhao}, X.~F. and {Tan}, Y. and {Nie}, J.~Y. and {Zhao}, H.~S. and {Zheng}, S.~J. and {Zheng}, Y.~G. and {Luo}, Q. and {Cai}, C. and {Li}, B. and {Xue}, W.~C. and {Bu}, Q.~C. and {Chang}, Z. and {Chen}, G. and {Chen}, L. and {Chen}, T.~X. and {Chen}, Y.~B. and {Chen}, Y.~P. and {Cui}, W. and {Cui}, W.~W. and {Deng}, J.~K. and {Dong}, Y.~W. and {Du}, Y.~Y. and {Fu}, M.~X. and {Gao}, G.~H. and {Gao}, H. and {Gao}, M. and {Gu}, Y.~D. and {Guan}, J. and {Guo}, C.~C. and {Han}, D.~W. and {Huang}, Y. and {Huo}, J. and {Jiang}, L.~H. and {Jiang}, W.~C. and {Jin}, J. and {Jin}, Y.~J. and {Kong}, L.~D. and {Li}, G. and {Li}, M.~S. and {Li}, W. and {Li}, X. and {Li}, X.~F. and {Li}, Y.~G. and {Li}, Z.~W. and {Liang}, X.~H. and {Liu}, B.~S. and {Liu}, G.~Q. and {Liu}, H.~W. and {Liu}, X.~J. and {Liu}, Y.~N. and {Lu}, B. and {Lu}, X.~F. and {Luo}, T. and {Ma}, X. and {Meng}, B. and {Ou}, G. and {Sai}, N. and {Shang}, R.~C. and {Song}, X.~Y. and {Sun}, L. and {Tao}, L. and {Wang}, C. and {Wang}, G.~F. and {Wang}, J. and {Wang}, W.~S. and {Wang}, Y.~S. and {Wen}, X.~Y. and {Wu}, B.~B. and {Wu}, B.~Y. and {Wu}, M. and {Xiao}, G.~C. and {Xu}, H. and {Yang}, J.~W. and {Yang}, S. and {Yang}, Y.~J. and {Yang}, Yi-Jung and {Yi}, Q.~B. and {Yin}, Q.~Q. and {You}, Y. and {Zhang}, A.~M. and {Zhang}, C.~M. and {Zhang}, F. and {Zhang}, H.~M. and {Zhang}, J. and {Zhang}, T. and {Zhang}, W. and {Zhang}, W.~C. and {Zhang}, W.~Z. and {Zhang}, Y. and {Zhang}, Yue and {Zhang}, Y.~F. and {Zhang}, Y.~J. and {Zhang}, Z. and {Zhang}, Zhi and {Zhang}, Z.~L. and {Zhou}, D.~K. and {Zhou}, J.~F. and {Zhu}, Y. and {Zhu}, Y.~X. and {Zhuang}, R.~L.},
        title = "{HXMT identification of a non-thermal X-ray burst from SGR J1935+2154 and with FRB 200428}",
      journal = {Nature Astronomy},
         year = 2021,
        month = apr,
       volume = {5},
        pages = {378-384},
          doi = {10.1038/s41550-021-01302-6},
archivePrefix = {arXiv},
       eprint = {2005.11071},
 primaryClass = {astro-ph.HE},
       adsurl = {https://ui.adsabs.harvard.edu/abs/2021NatAs...5..378L}
}

@ARTICLE{2021NatAs...5..401T,
       author = {{Tavani}, M. and {Casentini}, C. and {Ursi}, A. and {Verrecchia}, F. and {Addis}, A. and {Antonelli}, L.~A. and {Argan}, A. and {Barbiellini}, G. and {Baroncelli}, L. and {Bernardi}, G. and {Bianchi}, G. and {Bulgarelli}, A. and {Caraveo}, P. and {Cardillo}, M. and {Cattaneo}, P.~W. and {Chen}, A.~W. and {Costa}, E. and {Del Monte}, E. and {Di Cocco}, G. and {Di Persio}, G. and {Donnarumma}, I. and {Evangelista}, Y. and {Feroci}, M. and {Ferrari}, A. and {Fioretti}, V. and {Fuschino}, F. and {Galli}, M. and {Gianotti}, F. and {Giuliani}, A. and {Labanti}, C. and {Lazzarotto}, F. and {Lipari}, P. and {Longo}, F. and {Lucarelli}, F. and {Magro}, A. and {Marisaldi}, M. and {Mereghetti}, S. and {Morelli}, E. and {Morselli}, A. and {Naldi}, G. and {Pacciani}, L. and {Parmiggiani}, N. and {Paoletti}, F. and {Pellizzoni}, A. and {Perri}, M. and {Perotti}, F. and {Piano}, G. and {Picozza}, P. and {Pilia}, M. and {Pittori}, C. and {Puccetti}, S. and {Pupillo}, G. and {Rapisarda}, M. and {Rappoldi}, A. and {Rubini}, A. and {Setti}, G. and {Soffitta}, P. and {Trifoglio}, M. and {Trois}, A. and {Vercellone}, S. and {Vittorini}, V. and {Giommi}, P. and {D'Amico}, F.},
        title = "{An X-ray burst from a magnetar enlightening the mechanism of fast radio bursts}",
      journal = {Nature Astronomy},
         year = 2021,
        month = apr,
       volume = {5},
        pages = {401-407},
          doi = {10.1038/s41550-020-01276-x},
archivePrefix = {arXiv},
       eprint = {2005.12164},
 primaryClass = {astro-ph.HE},
       adsurl = {https://ui.adsabs.harvard.edu/abs/2021NatAs...5..401T}
}

@ARTICLE{2019PhR...821....1P,
       author = {{Platts}, E. and {Weltman}, A. and {Walters}, A. and {Tendulkar}, S.~P. and {Gordin}, J.~E.~B. and {Kandhai}, S.},
        title = "{A living theory catalogue for fast radio bursts}",
      journal = {\physrep},
         year = 2019,
        month = aug,
       volume = {821},
        pages = {1-27},
          doi = {10.1016/j.physrep.2019.06.003},
archivePrefix = {arXiv},
       eprint = {1810.05836},
 primaryClass = {astro-ph.HE},
       adsurl = {https://ui.adsabs.harvard.edu/abs/2019PhR...821....1P}
}

@ARTICLE{1980ApJ...236L.109L,
       author = {{Linscott}, I.~R. and {Erkes}, J.~W.},
        title = "{Discovery of millisecond radio bursts from M 87}",
      journal = {\apjl},
         year = 1980,
        month = mar,
       volume = {236},
        pages = {L109-L113},
          doi = {10.1086/183209},
       adsurl = {https://ui.adsabs.harvard.edu/abs/1980ApJ...236L.109L}
}

@ARTICLE{2021Univ....7..453C,
       author = {{Caleb}, Manisha and {Keane}, Evan},
        title = "{A Decade and a Half of Fast Radio Burst Observations}",
      journal = {Universe},
         year = 2021,
        month = nov,
       volume = {7},
       number = {11},
          eid = {453},
        pages = {453},
          doi = {10.3390/universe7110453},
       adsurl = {https://ui.adsabs.harvard.edu/abs/2021Univ....7..453C}
}

@ARTICLE{2023RvMP...95c5005Z,
       author = {{Zhang}, Bing},
        title = "{The physics of fast radio bursts}",
      journal = {Reviews of Modern Physics},
         year = 2023,
        month = jul,
       volume = {95},
       number = {3},
          eid = {035005},
        pages = {035005},
          doi = {10.1103/RevModPhys.95.035005},
archivePrefix = {arXiv},
       eprint = {2212.03972},
 primaryClass = {astro-ph.HE},
       adsurl = {https://ui.adsabs.harvard.edu/abs/2023RvMP...95c5005Z}
}

@ARTICLE{2024Ap&SS.369...59L,
       author = {{Lorimer}, Duncan R. and {McLaughlin}, Maura A. and {Bailes}, Matthew},
        title = "{The discovery and significance of fast radio bursts}",
      journal = {\apss},
         year = 2024,
        month = jun,
       volume = {369},
       number = {6},
          eid = {59},
        pages = {59},
          doi = {10.1007/s10509-024-04322-6},
archivePrefix = {arXiv},
       eprint = {2405.19106},
 primaryClass = {astro-ph.HE},
       adsurl = {https://ui.adsabs.harvard.edu/abs/2024Ap&SS.369...59L}
}

@ARTICLE{2023arXiv231016932G,
       author = {{Giri}, Utkarsh and {Andersen}, Bridget C. and {Chawla}, Pragya and {Curtin}, Alice P. and {Fonseca}, Emmanuel and {Kaspi}, Victoria M. and {Lin}, Hsiu-Hsien and {Masui}, Kiyoshi W. and {Sand}, Ketan R. and {Scholz}, Paul and {Abbott}, Thomas C. and {Dong}, Fengqiu Adam and {Gaensler}, B.~M. and {Leung}, Calvin and {Michilli}, Daniele and {Bhardwaj}, Mohit and {M{\"u}nchmeyer}, Moritz and {Pandhi}, Ayush and {Pearlman}, Aaron B. and {Pleunis}, Ziggy and {Rafiei-Ravandi}, Masoud and {Reda}, Alex and {Shin}, Kaitlyn and {Smith}, Kendrick and {Stairs}, Ingrid H. and {Stenning}, David C. and {Tendulkar}, Shriharsh P.},
        title = "{Comprehensive Bayesian analysis of FRB-like bursts from SGR 1935+2154 observed by CHIME/FRB}",
      journal = {arXiv e-prints},
         year = 2023,
        month = oct,
          eid = {arXiv:2310.16932},
        pages = {arXiv:2310.16932},
          doi = {10.48550/arXiv.2310.16932},
archivePrefix = {arXiv},
       eprint = {2310.16932},
 primaryClass = {astro-ph.HE},
       adsurl = {https://ui.adsabs.harvard.edu/abs/2023arXiv231016932G}
}

@ARTICLE{2020ApJ...898L..29M,
       author = {{Mereghetti}, S. and {Savchenko}, V. and {Ferrigno}, C. and {G{\"o}tz}, D. and {Rigoselli}, M. and {Tiengo}, A. and {Bazzano}, A. and {Bozzo}, E. and {Coleiro}, A. and {Courvoisier}, T.~J. -L. and {Doyle}, M. and {Goldwurm}, A. and {Hanlon}, L. and {Jourdain}, E. and {von Kienlin}, A. and {Lutovinov}, A. and {Martin-Carrillo}, A. and {Molkov}, S. and {Natalucci}, L. and {Onori}, F. and {Panessa}, F. and {Rodi}, J. and {Rodriguez}, J. and {S{\'a}nchez-Fern{\'a}ndez}, C. and {Sunyaev}, R. and {Ubertini}, P.},
        title = "{INTEGRAL Discovery of a Burst with Associated Radio Emission from the Magnetar SGR 1935+2154}",
      journal = {\apjl},
         year = 2020,
        month = aug,
       volume = {898},
       number = {2},
          eid = {L29},
        pages = {L29},
          doi = {10.3847/2041-8213/aba2cf},
archivePrefix = {arXiv},
       eprint = {2005.06335},
 primaryClass = {astro-ph.HE},
       adsurl = {https://ui.adsabs.harvard.edu/abs/2020ApJ...898L..29M}
}

@ARTICLE{2019PhRvD.100d3017D,
       author = {{de Souza}, Gibran H. and {Chirenti}, Cecilia},
        title = "{Torsional oscillations of magnetized neutron stars with mixed poloidal-toroidal fields}",
      journal = {\prd},
         year = 2019,
        month = aug,
       volume = {100},
       number = {4},
          eid = {043017},
        pages = {043017},
          doi = {10.1103/PhysRevD.100.043017},
archivePrefix = {arXiv},
       eprint = {1810.06628},
 primaryClass = {astro-ph.HE},
       adsurl = {https://ui.adsabs.harvard.edu/abs/2019PhRvD.100d3017D}
}

@ARTICLE{Beniamini+20,
	author = {{Beniamini}, Paz and {Wadiasingh}, Zorawar and {Metzger}, Brian D.},
	title = "{Periodicity in recurrent fast radio bursts and the origin of ultralong period magnetars}",
	journal = {\mnras},
	year = 2020,
	month = jun,
	volume = {496},
	number = {3},
	pages = {3390-3401},
	doi = {10.1093/mnras/staa1783},
	archivePrefix = {arXiv},
	eprint = {2003.12509},
	primaryClass = {astro-ph.HE},
	adsurl = {https://ui.adsabs.harvard.edu/abs/2020MNRAS.496.3390B}
}

@INPROCEEDINGS{Popov2010,
       author = {{Popov}, Sergey B. and {Postnov}, K.~A.},
        title = "{Hyperflares of SGRs as an engine for millisecond extragalactic radio bursts}",
    booktitle = {Evolution of Cosmic Objects through their Physical Activity},
         year = 2010,
       editor = {{Harutyunian}, H.~A. and {Mickaelian}, A.~M. and {Terzian}, Y.},
        month = nov,
        pages = {129-132},
          doi = {10.48550/arXiv.0710.2006},
archivePrefix = {arXiv},
       eprint = {0710.2006},
 primaryClass = {astro-ph},
       adsurl = {https://ui.adsabs.harvard.edu/abs/2010vaoa.conf..129P}
}

@ARTICLE{Wadiasingh2019,
       author = {{Wadiasingh}, Zorawar and {Timokhin}, Andrey},
        title = "{Repeating Fast Radio Bursts from Magnetars with Low Magnetospheric Twist}",
      journal = {\apj},
         year = 2019,
        month = jul,
       volume = {879},
       number = {1},
          eid = {4},
        pages = {4},
          doi = {10.3847/1538-4357/ab2240},
archivePrefix = {arXiv},
       eprint = {1904.12036},
 primaryClass = {astro-ph.HE},
       adsurl = {https://ui.adsabs.harvard.edu/abs/2019ApJ...879....4W}
}

@ARTICLE{Wadiasingh2020,
	author = {{Wadiasingh}, Zorawar and {Beniamini}, Paz and {Timokhin}, Andrey and {Baring}, Matthew G. and {van der Horst}, Alexander J. and {Harding}, Alice K. and {Kazanas}, Demosthenes},
	title = "{The Fast Radio Burst Luminosity Function and Death Line in the Low-twist Magnetar Model}",
	journal = {\apj},
	year = 2020,
	month = mar,
	volume = {891},
	number = {1},
	eid = {82},
	pages = {82},
	doi = {10.3847/1538-4357/ab6d69},
	archivePrefix = {arXiv},
	eprint = {1910.06979},
	primaryClass = {astro-ph.HE},
	adsurl = {https://ui.adsabs.harvard.edu/abs/2020ApJ...891...82W}
}

@article{PhysRevLett.77.4134,
  title = {Gravitational Waves and Pulsating Stars: What Can We Learn from Future Observations?},
  author = {Andersson, Nils and Kokkotas, Kostas D.},
  journal = {Phys. Rev. Lett.},
  volume = {77},
  issue = {20},
  pages = {4134--4137},
  numpages = {0},
  year = {1996},
  month = {Nov},
  publisher = {American Physical Society},
  doi = {10.1103/PhysRevLett.77.4134},
  url = {https://link.aps.org/doi/10.1103/PhysRevLett.77.4134}
}

@article{10.1046/j.1365-8711.1998.01840.x,
    author = {Andersson, Nils and Kokkotas, Kostas D.},
    title = {Towards gravitational wave asteroseismology},
    journal = {Monthly Notices of the Royal Astronomical Society},
    volume = {299},
    number = {4},
    pages = {1059-1068},
    year = {1998},
    month = {10},
    issn = {0035-8711},
    doi = {10.1046/j.1365-8711.1998.01840.x},
    url = {https://doi.org/10.1046/j.1365-8711.1998.01840.x},
    eprint = {https://academic.oup.com/mnras/article-pdf/299/4/1059/3869494/299-4-1059.pdf},
}

@article{PhysRevLett.120.172703,
  title = {Gravitational-Wave Constraints on the Neutron-Star-Matter Equation of State},
  author = {Annala, Eemeli and Gorda, Tyler and Kurkela, Aleksi and Vuorinen, Aleksi},
  journal = {Phys. Rev. Lett.},
  volume = {120},
  issue = {17},
  pages = {172703},
  numpages = {5},
  year = {2018},
  month = {Apr},
  publisher = {American Physical Society},
  doi = {10.1103/PhysRevLett.120.172703},
  url = {https://link.aps.org/doi/10.1103/PhysRevLett.120.172703}
}

@ARTICLE{2024NatAs...8..230K,
       author = {{Kramer}, Michael and {Liu}, Kuo and {Desvignes}, Gregory and {Karuppusamy}, Ramesh and {Stappers}, Ben W.},
        title = "{Quasi-periodic sub-pulse structure as a unifying feature for radio-emitting neutron stars}",
      journal = {Nature Astronomy},
         year = 2024,
        month = feb,
       volume = {8},
        pages = {230-240},
          doi = {10.1038/s41550-023-02125-3},
archivePrefix = {arXiv},
       eprint = {2311.13762},
 primaryClass = {astro-ph.HE},
       adsurl = {https://ui.adsabs.harvard.edu/abs/2024NatAs...8..230K}
}

@ARTICLE{2020Natur.587...54C,
       author = {{CHIME/FRB Collaboration} and {Andersen}, B.~C. and {Bandura}, K.~M. and {Bhardwaj}, M. and {Bij}, A. and {Boyce}, M.~M. and {Boyle}, P.~J. and {Brar}, C. and {Cassanelli}, T. and {Chawla}, P. and {Chen}, T. and {Cliche}, J.-F. and {Cook}, A. and {Cubranic}, D. and {Curtin}, A.~P. and {Denman}, N.~T. and {Dobbs}, M. and {Dong}, F.~Q. and {Fandino}, M. and {Fonseca}, E. and {Gaensler}, B.~M. and {Giri}, U. and {Good}, D.~C. and {Halpern}, M. and {Hill}, A.~S. and {Hinshaw}, G.~F. and {H{\"o}fer}, C. and {Josephy}, A. and {Kania}, J.~W. and {Kaspi}, V.~M. and {Landecker}, T.~L. and {Leung}, C. and {Li}, D.~Z. and {Lin}, H.-H. and {Masui}, K.~W. and {McKinven}, R. and {Mena-Parra}, J. and {Merryfield}, M. and {Meyers}, B.~W. and {Michilli}, D. and {Milutinovic}, N. and {Mirhosseini}, A. and {M{\"u}nchmeyer}, M. and {Naidu}, A. and {Newburgh}, L.~B. and {Ng}, C. and {Patel}, C. and {Pen}, U.-L. and {Pinsonneault-Marotte}, T. and {Pleunis}, Z. and {Quine}, B.~M. and {Rafiei-Ravandi}, M. and {Rahman}, M. and {Ransom}, S.~M. and {Renard}, A. and {Sanghavi}, P. and {Scholz}, P. and {Shaw}, J.~R. and {Shin}, K. and {Siegel}, S.~R. and {Singh}, S. and {Smegal}, R.~J. and {Smith}, K.~M. and {Stairs}, I.~H. and {Tan}, C.~M. and {Tendulkar}, S.~P. and {Tretyakov}, I. and {Vanderlinde}, K. and {Wang}, H. and {Wulf}, D. and {Zwaniga}, A.~V.},
        title = "{A bright millisecond-duration radio burst from a Galactic magnetar}",
      journal = {\nat},
         year = 2020,
        month = nov,
       volume = {587},
       number = {7832},
        pages = {54-58},
          doi = {10.1038/s41586-020-2863-y},
archivePrefix = {arXiv},
       eprint = {2005.10324},
 primaryClass = {astro-ph.HE},
       adsurl = {https://ui.adsabs.harvard.edu/abs/2020Natur.587...54C}
}

@ARTICLE{2024ATel16613....1B,
       author = {{Bhardwaj}, Mohit and {Kirichenko}, Aida and {Gil de Paz}, Armando},
        title = "{A redshift for the host galaxy of FRB 20240114A}",
      journal = {The Astronomer's Telegram},
         year = 2024,
        month = may,
       volume = {16613},
        pages = {1},
       adsurl = {https://ui.adsabs.harvard.edu/abs/2024ATel16613....1B}
}

@ARTICLE{2020Natur.587...59B,
       author = {{Bochenek}, C.~D. and {Ravi}, V. and {Belov}, K.~V. and {Hallinan}, G. and {Kocz}, J. and {Kulkarni}, S.~R. and {McKenna}, D.~L.},
        title = "{A fast radio burst associated with a Galactic magnetar}",
      journal = {\nat},
         year = 2020,
        month = nov,
       volume = {587},
       number = {7832},
        pages = {59-62},
          doi = {10.1038/s41586-020-2872-x},
archivePrefix = {arXiv},
       eprint = {2005.10828},
 primaryClass = {astro-ph.HE},
       adsurl = {https://ui.adsabs.harvard.edu/abs/2020Natur.587...59B}
}

@article{PhysRevC.85.035801,
  title = {Neutron conduction in the inner crust of a neutron star in the framework of the band theory of solids},
  author = {Chamel, N.},
  journal = {Phys. Rev. C},
  volume = {85},
  issue = {3},
  pages = {035801},
  numpages = {7},
  year = {2012},
  month = {Mar},
  publisher = {American Physical Society},
  doi = {10.1103/PhysRevC.85.035801},
  url = {https://link.aps.org/doi/10.1103/PhysRevC.85.035801}
}

@misc{chen2025hostgalaxyhyperactiverepeating,
      title={The Host Galaxy of the Hyperactive Repeating FRB 20240114A: Behind a Galaxy Cluster}, 
      author={Xiang-Lei Chen and Chao-Wei Tsai and Di Li and Pei Wang and Yi Feng and Jun-Shuo Zhang and Guo-Dong Li and Yong-Kun Zhang and Lu-Lu Bao and Mai Liao and Lu-Dan Zhang and Pei Zuo and Dong-Wei Bao and Chen-Hui Niu and Rui Luo and Wei-Wei Zhu and Hu Zou and Sui-Jian Xue and Bing Zhang},
      year={2025},
      eprint={2502.05587},
      archivePrefix={arXiv},
      primaryClass={astro-ph.GA},
      url={https://arxiv.org/abs/2502.05587}, 
}

@ARTICLE{2024ApJ...974..295D,
       author = {{Dittmann}, Alexander J. and {Miller}, M. Coleman and {Lamb}, Frederick K. and {Holt}, Isiah M. and {Chirenti}, Cecilia and {Wolff}, Michael T. and {Bogdanov}, Slavko and {Guillot}, Sebastien and {Ho}, Wynn C.~G. and {Morsink}, Sharon M. and {Arzoumanian}, Zaven and {Gendreau}, Keith C.},
        title = "{A More Precise Measurement of the Radius of PSR J0740+6620 Using Updated NICER Data}",
      journal = {\apj},
         year = 2024,
        month = oct,
       volume = {974},
       number = {2},
          eid = {295},
        pages = {295},
          doi = {10.3847/1538-4357/ad5f1e},
archivePrefix = {arXiv},
       eprint = {2406.14467},
 primaryClass = {astro-ph.HE},
       adsurl = {https://ui.adsabs.harvard.edu/abs/2024ApJ...974..295D}
}

@ARTICLE{2010Natur.467.1081D,
       author = {{Demorest}, P.~B. and {Pennucci}, T. and {Ransom}, S.~M. and {Roberts}, M.~S.~E. and {Hessels}, J.~W.~T.},
        title = "{A two-solar-mass neutron star measured using Shapiro delay}",
      journal = {\nat},
         year = 2010,
        month = oct,
       volume = {467},
       number = {7319},
        pages = {1081-1083},
          doi = {10.1038/nature09466},
archivePrefix = {arXiv},
       eprint = {1010.5788},
 primaryClass = {astro-ph.HE},
       adsurl = {https://ui.adsabs.harvard.edu/abs/2010Natur.467.1081D}
}

@article{PhysRevD.88.044052,
  title = {Gravitational wave asteroseismology of fast rotating neutron stars with realistic equations of state},
  author = {Doneva, Daniela D. and Gaertig, Erich and Kokkotas, Kostas D. and Kr\"uger, Christian},
  journal = {Phys. Rev. D},
  volume = {88},
  issue = {4},
  pages = {044052},
  numpages = {18},
  year = {2013},
  month = {Aug},
  publisher = {American Physical Society},
  doi = {10.1103/PhysRevD.88.044052},
  url = {https://link.aps.org/doi/10.1103/PhysRevD.88.044052}
}

@article{10.1093/mnras/sts721,
    author = {Gabler, Michael and Cerd\'{a}-Dur\'{a}n, Pablo and Font, Jos\{e} A. and M\"{u}ller, Ewald and Stergioulas, Nikolaos}

@article{10.1093/mnras/sty445,
    author = {Gabler, Michael and Cerd\'{a}-Dur\'{a}n, Pablo and Stergioulas, Nikolaos and Font, Jos\'{e} A and M\"{u}ller, Ewald},
    title = {Constraining properties of high-density matter in neutron stars with magneto-elastic oscillations},
    journal = {Monthly Notices of the Royal Astronomical Society},
    volume = {476},
    number = {3},
    pages = {4199-4212},
    year = {2018},
    month = {03},
    issn = {0035-8711},
    doi = {10.1093/mnras/sty445},
    url = {https://doi.org/10.1093/mnras/sty445},
    eprint = {https://academic.oup.com/mnras/article-pdf/476/3/4199/24541867/sty445.pdf},
}


\end{document}